\documentclass[aps,prd,groupedaddress]{revtex4}
\usepackage{graphicx}
\usepackage{epsfig}
\usepackage{amsfonts}
\usepackage{amssymb}
\usepackage{epsf}
\usepackage{color}
\newcommand{\insertplot}[5]{\begin{figure}
 \hfill\hbox to 0.05in{\vbox to #5in{\vfill
 \inputplot{#1}{#4}{#5}}\hfill}
 \hfill\vspace{-.1in}
 \caption{#2}\label{#3}
 \end{figure}}
\newcommand{\inputplot}[3]{
 \special{ps: plotfile #1}
}

\newcounter{fig}

\begin{document}

\title{Spontaneously vectorized Einstein-Gauss-Bonnet black holes}
\author{Simon Barton}
\email[]{barton@ph1.uni-koeln.de}
\affiliation{Faculty of Mathematics and Natural Sciences, University of Cologne, D-50923 Cologne, Germany}
\author{Betti Hartmann}
\email[]{betti.hartmann@uni-oldenburg.de}
\affiliation{Instituto de F\'isica de S\~ao Carlos, Universidade de S\~ao Paulo, 
S\~ao Carlos, S\~ao Paulo, 13560-970, Brazil\\ 
Department of Physics and Earth Sciences, Jacobs University Bremen, 28759 Bremen, Germany\\
Institute of Physics, University of Oldenburg, D-26111 Oldenburg, Germany\\
Department of Mathematics, University College London, Gower Street, London, WC1E 6BT, UK
}
\author{Burkhard Kleihaus}
\email[]{b.kleihaus@uni-oldenburg.de}
\affiliation{Institute of Physics, University of Oldenburg, D-26111 Oldenburg, Germany}
\author{Jutta Kunz}
\email[]{jutta.kunz@uni-oldenburg.de}
\affiliation{Institute of Physics, University of Oldenburg, D-26111 Oldenburg, Germany}

\date{\today}

\begin{abstract}
We construct spontaneously vectorized black holes where a real vector field is
coupled to the Gauss-Bonnet invariant.
We employ three coupling functions for the vector field,
and determine the respective domains of existence of the vectorized black holes.
These domains of existence are bounded by the marginally stable Schwarzschild black holes
and the critical vectorized black holes.
We also address the effects of a mass term. 
For a given black hole mass the horizon radius is smaller
for the vectorized black holes than for the Schwarzschild black holes.
Since the vector field vanishes at the horizon,
there is no contribution from the Gauss-Bonnet term to the entropy
of the vectorized black holes.
\end{abstract}

\maketitle

\section{Introduction}
Black holes in General Relativity (GR) satisfy uniqueness theorems 
\cite{Chrusciel:2012jk}.
The Schwarzschild and Kerr black holes represent
the static, respectively  stationary rotating,  
black hole solutions of the Einstein equations in vacuum.
When a real scalar field is admitted, the Schwarzschild and Kerr black holes
remain the only black hole solutions: 
Schwarzschild and Kerr black holes in GR carry no real scalar hair
(see e.g., \cite{Herdeiro:2015waa}).
Inclusion of a massless vector field, however, leads to the Reissner-Nordstr\"om
and Kerr-Newman black holes of Einstein-Maxwell theory, for which
again uniqueness theorems hold \cite{Chrusciel:2012jk}.

When going beyond GR black holes may carry real scalar fields.
For instance, GR may be amended by higher curvature terms,
that are coupled to a scalar field. A particular higher curvature term
is the Gauss-Bonnet (GB) invariant, whose presence is well-motivated
from quantum gravity considerations 
\cite{Zwiebach:1985uq,Gross:1986mw,Metsaev:1987zx}.
Moreover, the resulting Einstein-scalar-Gauss-Bonnet (EsGB) theories
possess second order field equations and thus avoid Ostrogradski instability and ghosts 
\cite{Horndeski:1974wa,Charmousis:2011bf,Kobayashi:2011nu}.

The coupling of the scalar field to the GB invariant represents a
non-minimal coupling, where the coupling function can be chosen freely.
A string theory motivated dilatonic coupling function 
leads to black holes, which are always scalarized
\cite{Kanti:1995vq,Torii:1996yi,Guo:2008hf,Pani:2009wy,Pani:2011gy,Kleihaus:2011tg,Ayzenberg:2013wua,Ayzenberg:2014aka,Maselli:2015tta,Kleihaus:2014lba,Kleihaus:2015aje,Blazquez-Salcedo:2016enn,Blazquez-Salcedo:2017txk}
(see also \cite{Sotiriou:2013qea,Sotiriou:2014pfa}).
In this case, the scalar field equation always has a non-vanishing source term,
since the derivative of the coupling function with respect to the scalar field is always finite.
Therefore the Schwarzschild and Kerr black holes are no longer solutions
of the coupled set of EsGB equations.

However, the coupling function can also be chosen to allow
for the Schwarzschild and Kerr black holes to be solutions
of the coupled set of EsGB equations.
In this case the derivative of the coupling function with respect to the scalar field should
vanish for some value of the scalar field,
such that the scalar field can be chosen 
to have this constant value throughout.
While the Schwarzschild and Kerr black holes remain solutions
of the EsGB equations, they do not remain the only solutions,
since spontaneously scalarized black hole solutions arise as well
\cite{Antoniou:2017acq,Doneva:2017bvd,Silva:2017uqg,Antoniou:2017hxj,Blazquez-Salcedo:2018jnn,Doneva:2018rou,Minamitsuji:2018xde,Silva:2018qhn,Brihaye:2018grv,
Bakopoulos:2018nui,Doneva:2019vuh,Cunha:2019dwb,
Macedo:2019sem,Myung:2019wvb,Hod:2019pmb,Bakopoulos:2019tvc, Collodel:2019kkx,Bakopoulos:2020dfg,
Blazquez-Salcedo:2020rhf,Blazquez-Salcedo:2020caw,Herdeiro:2020wei,Berti:2020kgk}
(see also \cite{Dima:2020yac,Hod:2020jjy,Doneva:2020nbb,Doneva:2020kfv,Doneva:2021dqn}).

In this case, Schwarzschild and Kerr black holes remain solutions of the EsGB equations
independent of the value of the GB coupling constant. However, they lose their
stability when scalarization sets in. 
In particular, the GB invariant leads to a tachyonic instability, since it features
in the scalar field equation like an effective mass. At a certain threshold value
of the GB coupling constant, the GR black holes then develop a zero mode,
where a branch of scalarized EsGB black holes emerges.
The first zero mode gives rise to the fundamental branch of
scalarized black holes, while the next zero modes give rise
to radially and angularly excited scalarized black holes.
Depending on the coupling function, the fundamental scalarized mode
may be (at least in part) stable or unstable
\cite{Antoniou:2017acq,Doneva:2017bvd,Silva:2017uqg,Antoniou:2017hxj,Blazquez-Salcedo:2018jnn,Doneva:2018rou,Minamitsuji:2018xde,Silva:2018qhn,Brihaye:2018grv,
Bakopoulos:2018nui,Doneva:2019vuh,Cunha:2019dwb,
Macedo:2019sem,Myung:2019wvb,Hod:2019pmb,Bakopoulos:2019tvc, Collodel:2019kkx,Bakopoulos:2020dfg,
Blazquez-Salcedo:2020rhf,Blazquez-Salcedo:2020caw,Herdeiro:2020wei,Berti:2020kgk}.

Spontaneous scalarization of Reissner-Nordstr\"om (RN)
and Kerr-Newman black holes can be achieved in GR, when the scalar field is non-minimally coupled
to the Maxwell invariant with an appropriate coupling function
\cite{Herdeiro:2018wub,Myung:2018vug,Boskovic:2018lkj,Myung:2018jvi,Fernandes:2019rez,Brihaye:2019kvj,Myung:2019oua,Astefanesei:2019pfq,Konoplya:2019goy,Fernandes:2019kmh,Zou:2019bpt,Brihaye:2019gla,Astefanesei:2019qsg}.
Here the finite value of the Maxwell invariant of a charged black hole provides the effective mass term
necessary for the tachyonic instability of the GR black holes.
However, for particular choices of coupling functions,
also scalarized black hole can arise and coexist with the GR black holes  
without a tachyonic instability of the GR black holes ever occurring
\cite{Astefanesei:2019pfq,Blazquez-Salcedo:2020nhs,Blazquez-Salcedo:2020jee,Blazquez-Salcedo:2020crd}.

However, besides spontaneous scalarization of black holes 
also spontaneous vectorization of black holes may occur,
as argued vigorously by Ramazano\u{g}lu
\cite{Ramazanoglu:2017xbl,Ramazanoglu:2018tig,Ramazanoglu:2019gbz,Ramazanoglu:2019jrr}
(see also \cite{Fan:2016jnz}).
In this case a vector field has to be coupled to an invariant with a suitable coupling function.
The black holes of GR then remain solutions of the generalized set of field equations,
but succumb to a tachyonic instability induced by the contribution from the invariant
in the vector field equation acting as an effective mass.
Recently such spontaneously vectorized black hole solutions have been obtained 
in GR, where an additional vector field has been non-minimally coupled
to the Maxwell invariant with an appropriate coupling function
\cite{Oliveira:2020dru}.

Here we construct and investigate spontaneously vectorized black hole solutions of 
Einstein-vector-Gauss-Bonnet (EvGB) theories.
We employ several coupling functions, which all satisfy the criteria for
spontaneous vectorization: they are functions of the vector field squared,
$A_\mu A^\mu$, thus for a vanishing vector field the coupling functions
vanish, allowing the Schwarzschild black hole solutions to remain 
solutions of the EvGB equations.
Since the GB term enters the vector field equations like an effective mass term,
a tachyonic instability of the GR black holes results,
giving rise to branches of vectorized black holes.

We have organized the paper as follows:
Section II describes the theoretical setting
with the action, the equations of motion,
and the boundary conditions,
and we define the physical properties.
Section III contains our physical results,
together with a brief description of the numerics.
Here we discuss the solutions, the domain of existence
and the physical properties of the black holes.
We give our conclusions in section IV.

\section{Theoretical setting}

\subsection{Action and equations of motion}

We consider the effective action for EvGB theories 
\begin{eqnarray}  
S=\frac{1}{16 \pi}\int \left[R - F_{\mu\nu}F^{\mu\nu}  -V(A_\mu A^\mu) 
 + F(A_\mu A^\mu) R^2_{\rm GB}
 \right] \sqrt{-g} d^4x  \ ,
\label{act}
\end{eqnarray} 
where $R$ is the curvature scalar,
and $F_{\mu\nu}$ denotes the field strength tensor
of the real vector field $A_\mu$ with potential $V(A_\mu A^\mu)$.
The vector field is coupled with some coupling function $F(A_\mu A^\mu)$ to 
the Gauss-Bonnet term
\begin{eqnarray} 
R^2_{\rm GB} = R_{\mu\nu\rho\sigma} R^{\mu\nu\rho\sigma}
- 4 R_{\mu\nu} R^{\mu\nu} + R^2 \ . 
\end{eqnarray} 
For the coupling function $F(A_\mu A^\mu)$ we make the following choices
\begin{eqnarray}
F(A_\mu A^\mu) = \left\{\begin{array}{ccc}
\lambda \left( 1 - e^{  - \beta A_\mu A^\mu} \right)  & \hspace{1.0cm} & (i)\\
\lambda \left( e^{   \beta A_\mu A^\mu} -1 \right)  & \hspace{1.0cm} & (ii)\\
\lambda A_\mu A^\mu  & \hspace{1.0cm} & (iii)
\end{array} \right. 
\end{eqnarray}
with coupling constants $\lambda$ and $\beta$.
When the vector field vanishes, $A_\mu=0$, all three coupling functions reduce to zero.
The potential $V(A_\mu A^\mu)$ 
\begin{equation}
V(A_\mu A^\mu) = 2 m_A^2 A_\mu A^\mu  -2 \alpha \left(A_\mu A^\mu\right)^2
\end{equation}
has a mass term with vector field mass $m_A$
and a self-interaction with coupling constant $\alpha$.
We here mostly focus on $\alpha=0$.
While the Gauss-Bonnet invariant $R^2_{\rm GB}$ itself is topological in four dimensions, 
its coupling to the vector field $A_\mu$ 
by means of the coupling function $F(A_\mu A^\mu)$ leads to
significant contributions to the equations of motion.

The coupled set of field equations follows from the variational principle.
Variation of the action (\ref{act}) with respect to 
the vector field and the metric yields the Proca equation and the Einstein equations
\begin{equation}
\nabla_\mu F^{\mu\nu} =  \frac{1}{2} \frac{dV(A_\mu A^\mu)}{d (A_\mu A^\mu)} A^\nu -
\frac{1}{2} \frac{dF(A_\mu A^\mu)}{d (A_\mu A^\mu)} R^2_{\rm GB} A^\nu \ ,
\label{scleq}
\end{equation}
\begin{equation}
G_{\mu\nu}  = \frac{1}{2}T^{({\rm eff})}_{\mu\nu} \ , 
\label{Einsteq}
\end{equation}
where $G_{\mu\nu}$ is the Einstein tensor and $T^{({\rm eff})}_{\mu\nu}$
denotes the effective stress-energy tensor	
\begin{equation}
T^{({\rm eff})}_{\mu\nu} = T^{(A)}_{\mu\nu} -2 T^{(GB)}_{\mu\nu} \ ,
\label{teff}
\end{equation}
which consists of a contribution from the vector field
\begin{equation}
T^{(A)}_{\mu\nu} = 4 F_\mu^{\phantom{\mu}\lambda} F_{\nu\lambda} 
                    +2\frac{dV(A_\lambda A^\lambda)}{d (A_\lambda A^\lambda)} A_\mu A_\nu
		    -g_{\mu\nu} \left(F_{\rho\lambda}F^{\rho\lambda} + V(A_\lambda A^\lambda)\right)\ , 
\label{tphi}
\end{equation}
and a contribution from the GB term $R^2_{\rm GB}$
\begin{equation}
T^{(GB)}_{\mu\nu} =
\frac{1}{2}\left(g_{\rho\mu} g_{\lambda\nu}+g_{\lambda\mu} g_{\rho\nu}\right)
\eta^{\kappa\lambda\alpha\beta}\tilde{R}^{\rho\gamma}_{\phantom{\rho\gamma}\alpha\beta}
\nabla_\gamma \nabla_\kappa F(A_\mu A^\mu) + 
R^2_{\rm GB}\frac{d F(A_\sigma A^\sigma)}{d (A_\sigma A^\sigma)} A_\mu A_\nu\ ,
\label{teffi}
\end{equation}
where
$\tilde{R}^{\rho\gamma}_{\phantom{\rho\gamma}\alpha\beta}=\eta^{\rho\gamma\sigma\tau}
R_{\sigma\tau\alpha\beta}$ and $\eta^{\rho\gamma\sigma\tau}= 
\epsilon^{\rho\gamma\sigma\tau}/\sqrt{-g}$. Note that the last term results from the 
dependence of the coupling function on the metric.

To obtain static, spherically symmetric black holes we employ isotropic coordinates
for the line element
\begin{equation}
ds^2 = -F_0 dt^2 +e^{f_1}\left[dr^2 
+r^2\left( d\theta^2+\sin^2\theta d\varphi^2\right) \right]\ ,
\label{met}
\end{equation}
and we assume for the vector field the form
\begin{equation}
A_\mu dx^\mu = A_t dt \ .
\label{vec}
\end{equation}
All three functions,
the two metric functions $F_0$ and $f_1$ and the vector field function $A_t$,
depend only on the radial coordinate $r$. 

When we insert the above ansatz (\ref{met})-(\ref{vec}) for the metric and the vector field into
the set of EvGB equations
we obtain five coupled, nonlinear ordinary differential equations (ODEs).
However, these are not independent,
and one ODE can be treated as a constraint.
This leaves us with three second order ODEs.

Inspection of the field equations reveals an invariance under the scaling transformation
\begin{equation}
r \to \chi r \ , \ \ \    t \to \chi t \ , \ \ \  F \to \chi^2 F\  , \ \ \ V \to V/\chi^2 \ ,
\ \ \  \chi > 0 \ .
\label{scalinvar}
\end{equation}

\subsection{Black hole properties}

We are looking for vectorized black holes with a regular horizon.
Inspecting the equations of motion for the functions,
and performing an expansion at the horizon leads to
\begin{eqnarray}
F_0(r) &=& F_{02} \, \left( \frac{r- r_{\rm H}}{r_{\rm H}} \right)^2 + O  \left( \frac{r- r_{\rm H}}{r_{\rm H}} \right)^3 ,
\label{f0_H}\\
f_1(r) &=& f_1(r_{\rm H}) + O  \left( \frac{r- r_{\rm H}}{r_{\rm H}} \right) ,
\label{f1_H}\\
A_t(r) &=& A_{t2} \left( \frac{r- r_{\rm H}}{r_{\rm H}} \right)^2  + O \left( \frac{r- r_{\rm H}}{r_{\rm H}} \right)^3 ,
\label{At_H}
\end{eqnarray}
with constants $F_{02}$, $ f_1(r_{\rm H})$, and $A_{t1}$.
Thus at the horizon the metric function $F_0$ vanishes, while $f_1$ is finite. Interestingly,
also the vector field function $A_t$ vanishes at the horizon, but $A^t(r=r_{\rm H}) = A_{t2}/F_{02}$ is finite.

To address the physical properties of the vectorized black holes at the horizon we note that
the metric of a spatial cross-section of the horizon is
\begin{eqnarray}
\label{horizon-metric}
d\Sigma^2_{\rm H}=h_{ij} dx^i dx^j=
r_{\rm H}^2e^{f_1(r_{\rm H})}
\left( d\theta^2 + \sin^2\theta d\varphi^2\right) .
\end{eqnarray}
The horizon area of the black holes is then given by
 \begin{eqnarray}
\label{AH}
A_{\rm H}=4\pi r_{\rm H}^2  e^{f_1(r_{\rm H})} .
\end{eqnarray}
In GR the entropy is simply a quarter of the horizon area \cite{Wald:1984rg},
but this may be no longer the case in the presence of a GB term. 
In the case of scalarized black holes 
the entropy black holes acquires an additional contribution due to the coupling to the GB term
\cite{Lee:1990nz,Wald:1993nt,Iyer:1994ys,Hajian:2015xlp,Ghodrati:2016vvf,Hajian:2020dcq}.
For vectorized black holes an analogous additional term arises,
and the entropy can be expressed as the following integral over the horizon
\begin{eqnarray}
\label{S-Noether} 
S=\frac{1}{4}\int_{\Sigma_{\rm H}} d^{2}x 
\sqrt{h}\left[1+ 2 F(A_\mu A^\mu) \tilde R \right],
\end{eqnarray} 
where $h$ is the determinant of the induced metric on the horizon, Eq.~(\ref{horizon-metric}),
and $\tilde R$ is the horizon curvature.
Since, however, the vector field function $A_t$ vanishes at the horizon,
also the chosen coupling functions (i)-(iii) vanish at the horizon.
Therefore we obtain no contribution from the GB term to the entropy,
and the entropy remains equal to a quarter of the horizon area,
\begin{equation}
S= \frac{A_{\rm H}}{4} .
\label{entropy}
\end{equation}

The Killing vector field $\chi= \partial_t$ determines the surface gravity $\kappa$ \cite{Wald:1984rg},
where $\kappa^2=-\frac{1}{2}(\nabla_a \chi_b)(\nabla^a \chi^b)|_{r_{\rm H}}$,
yielding the Hawking temperature $T_{\rm H}={\kappa}/({2\pi})$
 \begin{eqnarray}
\label{TH}
T_{\rm H}=\frac{1}{2 \pi r_{\rm H}} \sqrt{ F_{02}} e^{-f_1(r_{\rm H})/2}   .
 \end{eqnarray}

We require the black hole solutions to be asymptotically flat. From the expansion at radial infinity 
\begin{eqnarray}
\label{asym1}
g_{tt} =-1+\frac{2M}{r}+\dots, \\
g_{rr} =1+\frac{2M}{r}+\dots, 
\label{asym3}\\
A_t = \frac{\tilde{Q}}{r}e^{-m_A r} + \dots ,
\label{asym2}
\end{eqnarray}  
with constants $M$ and $\tilde{Q}$,
we determine the asymptotic boundary conditions for the functions
\begin{equation}
F_0(\infty) =1 , \ \ \
f_1(\infty)=0 , \ \ \
A_t(\infty) = 0 \ .
\label{BC2}
\end{equation}
The constant $M$ in the expansion 
corresponds to the total mass of the black hole solutions.
This value agrees with the Komar mass,
when the Komar integral is evaluated at spatial infinity.
When the Komar integral is evaluated at the horizon, 
the horizon mass $M_{\rm H}$ is obtained.
For Schwarzschild black holes the horizon mass
$M_{\rm H}$ is identical to the total mass $M$.
For vectorized black holes this is no longer the case,
since the total mass $M$ receives a contribution from the bulk.

We define a vector charge $Q$ by the integral expression
\begin{equation}
Q =\frac{1}{4\pi} \int{\left. \sqrt{-g}F^{rt}\right|_{r\to\infty}} d\theta d\varphi 
= Q_{\rm H} + \frac{1}{4\pi}\int_{r> r_{\rm H}}\sqrt{-g} j^t d^3 x \ 
               \label{Qint}
\end{equation}
with the time component of the current density $j^\nu=\nabla_\mu F^{\mu \nu}$, 
and the horizon charge $Q_{\rm H}$
\begin{equation}
Q_{\rm H} = \frac{1}{4\pi} \int{\left. \sqrt{-g}F^{rt}\right|_{r=r_{\rm H}}} d\theta d\varphi \ .
\label{horzcharge}
\end{equation}
In the case of a massless vector field the vector charge $Q$
coincides with the constant $\tilde{Q}$, Eq.~(\ref{asym2}), 
whereas in the case of a massive vector field the charge vanishes at radial infinity, $Q=0$.
The horizon charge $Q_{\rm H}$, on the other hand, remains finite for massless and massive vector fields.

\section{Results}

\subsection{Numerics}

In order to solve the set of coupled Einstein and vector field equations numerically we 
introduce the radial coordinate
\begin{equation}
x = 1- \frac{r_{\rm H}}{r} \ ,
\label{x_com}
\end{equation}
to compactify the domain of integration, $0 \le x \le 1$.

The expansions close to the horizon, Eqs.~(\ref{f0_H}) and (\ref{At_H})
suggest to factorize the double-zeros of the functions $F_0$ and $A_t$,
\begin{equation}
F_0(x) = x^2 f_0(x) \ , \ \ \ A_t(x) = x^2  b(x) \ .
\label{newAn}
\end{equation}
Expansion of the Einstein and vector field equations close to $x=0$ then yields 
the boundary conditions at the horizon ($x=0$)
\begin{equation}
f'_0(0)-f_0(0) = 0 \ , \ \ \ 
f'_1(0)        = 2 \ , \ \ \ 
b'_0(0)-b(0) = 0 \ , 
\label{bc_horz}
\end{equation}
whereas the boundary conditions in the asymptotic region ($x=1$) are obtained from Eqs.~(\ref{BC2}),
\begin{equation}
f_0(1) = 1 \ , \ \ \ 
f_1(1) = 0 \ , \ \ \ 
b(1) = 0 \ . 
\label{bc_inf}
\end{equation}

We then employ the professional solver COLSYS \cite{Ascher:1979iha}.
COLSYS uses a collocation method to solve systems of boundary-value ODEs 
with the help of a damped Newton method of quasi-linearization
and an adaptive mesh selection procedure.
Starting from an initial guess, the iteration process then proceeds with successively
refined grids until a specified accuracy of the functions is reached.
When calculating the solutions we fix the isotropic horizon coordinate $r_{\rm H}=1$, 
and thus break the scaling invariance, Eqs.~(\ref{scalinvar}).

\subsection{Solutions}

\begin{figure}[t!]
\begin{center}
(a)\includegraphics[width=.45\textwidth, angle =0]{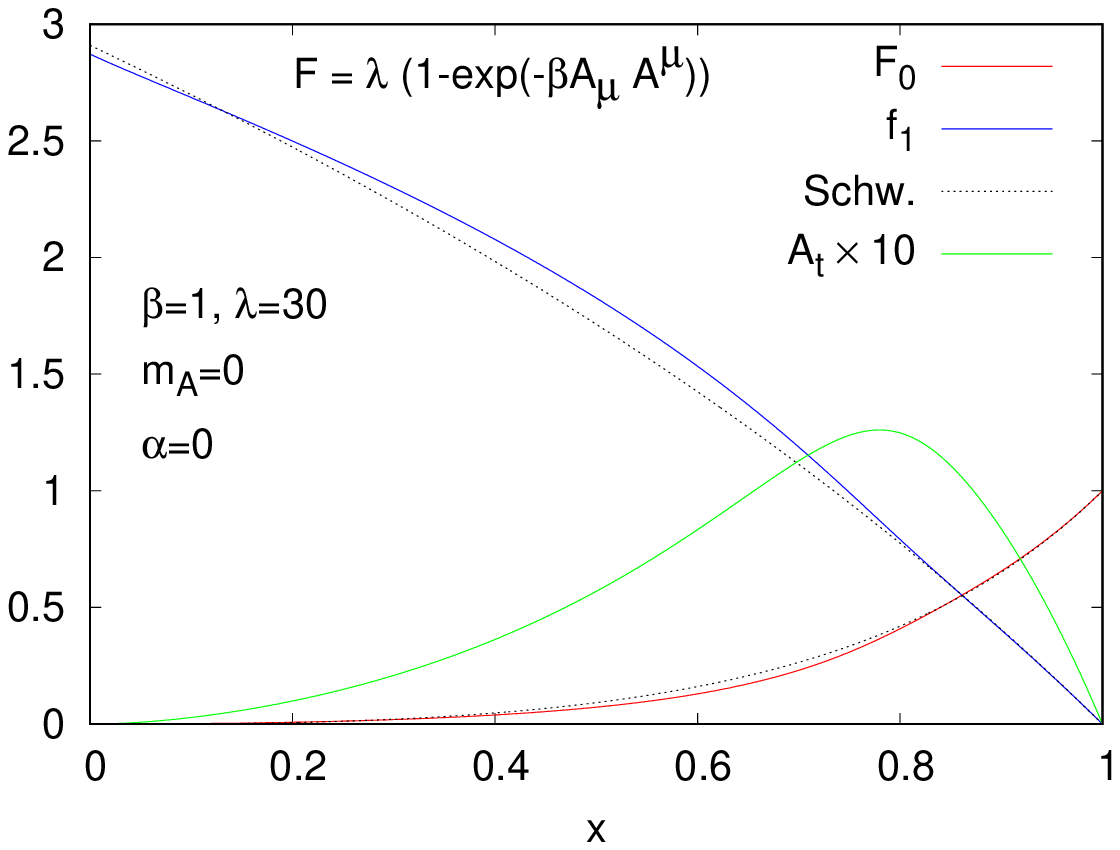}
(b)\includegraphics[width=.45\textwidth, angle =0]{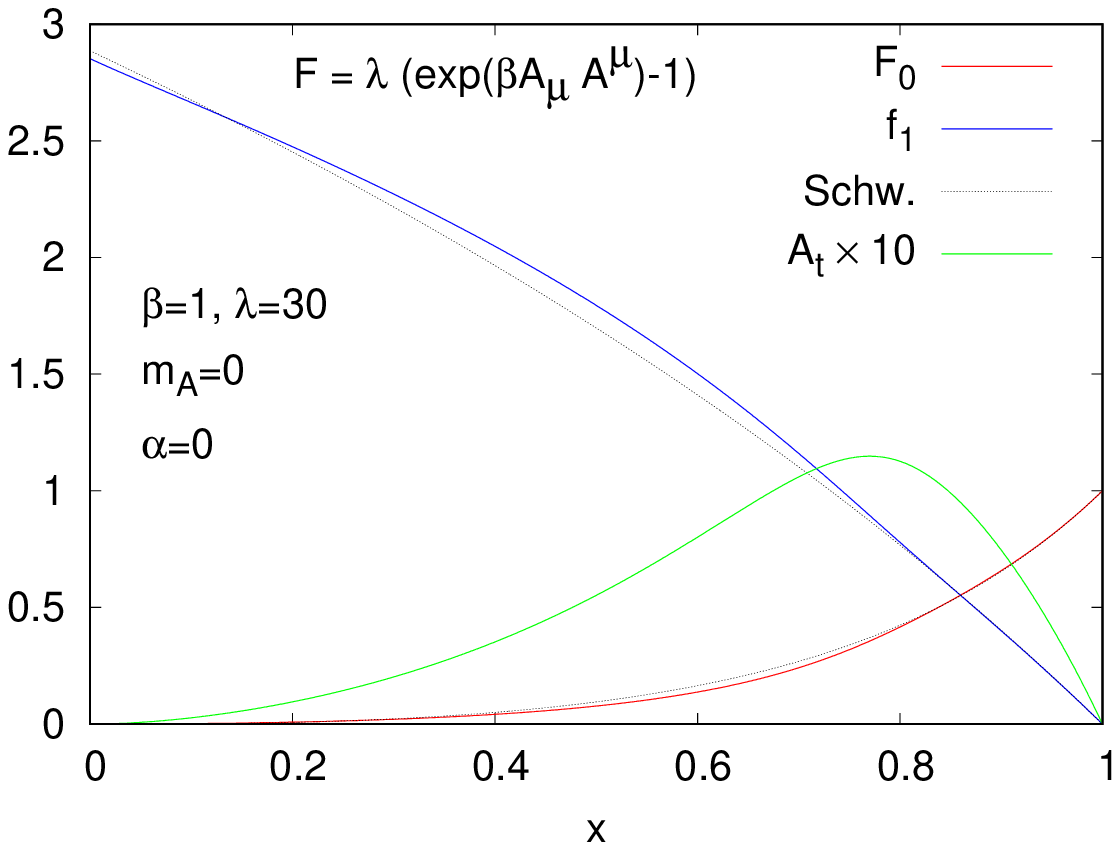}
\\
(c)\includegraphics[width=.45\textwidth, angle =0]{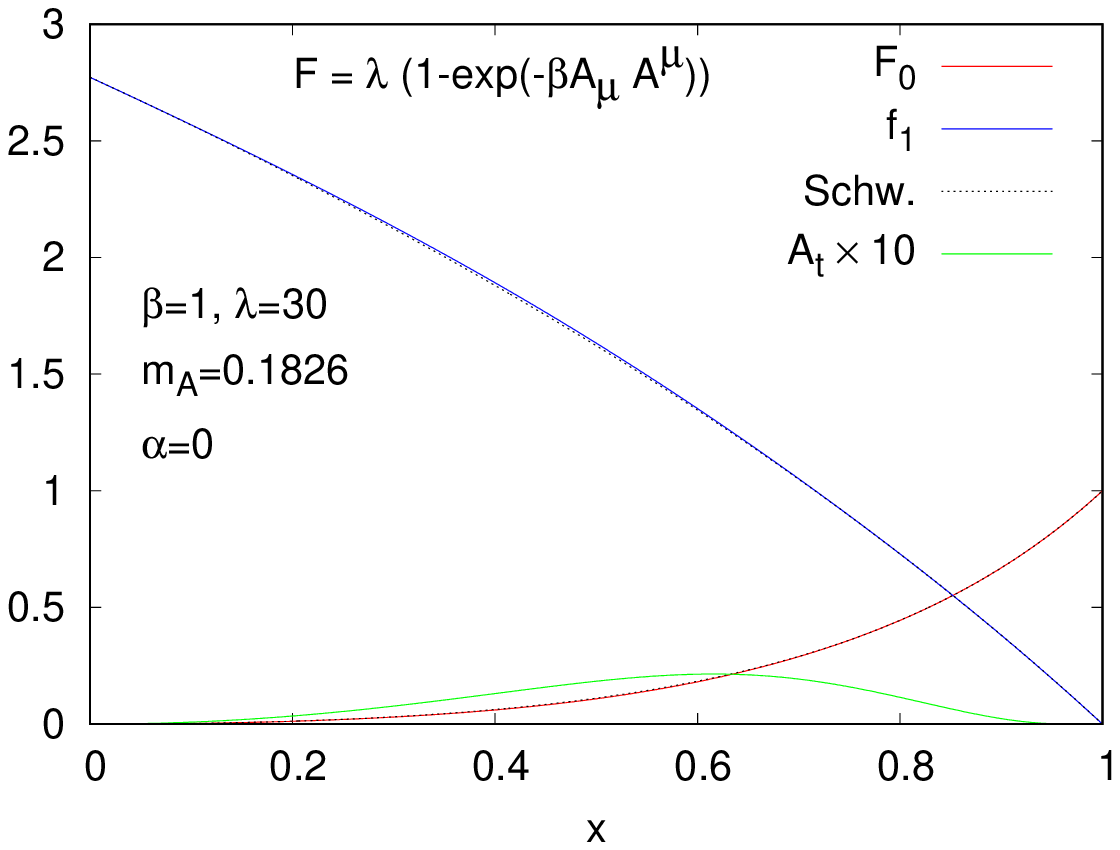}
(d)\includegraphics[width=.45\textwidth, angle =0]{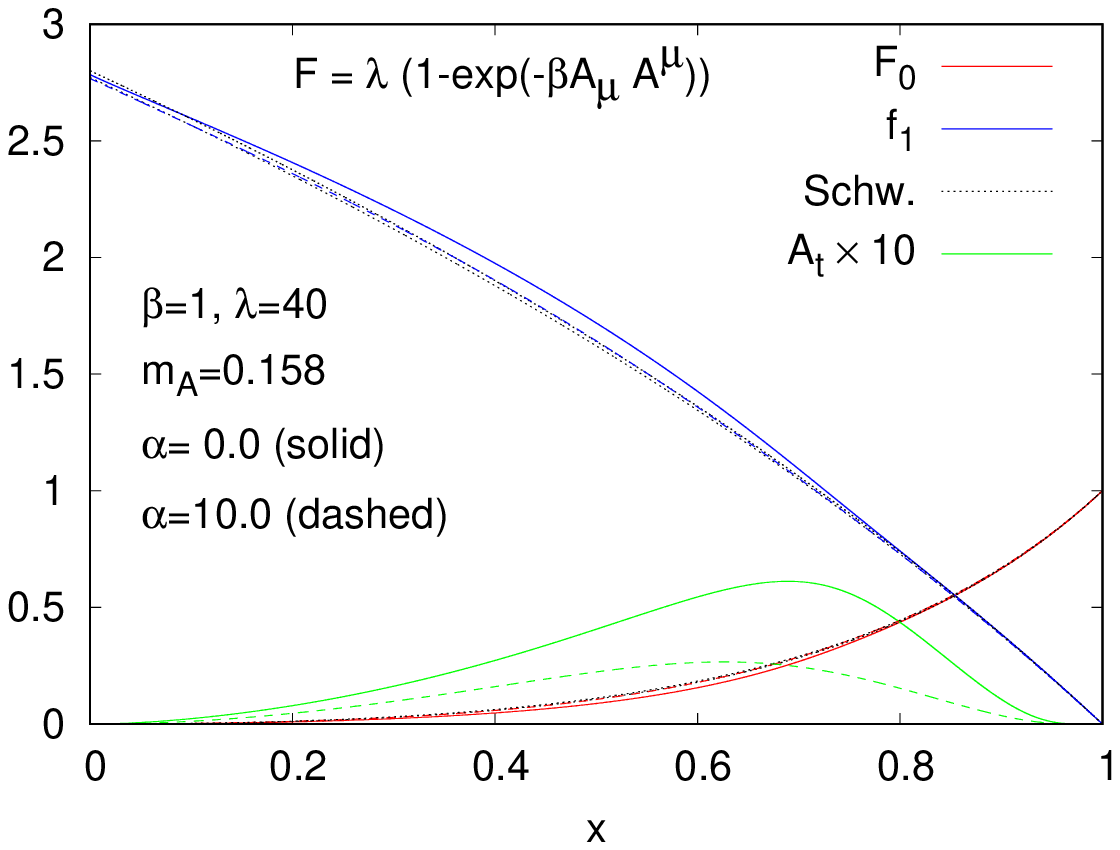}
\end{center}
\caption{Examples of vectorized black hole solutions:
metric functions $F_0$, $f_1$, and vector field function $A_t$
vs compactified radial coordinate $x$ for
(a) coupling function (i) and parameters $\lambda=30$, $\beta=1$, $m_A=0$, $\alpha=0$;
(b) coupling function (ii) and parameters $\lambda=30$, $\beta=1$, $m_A=0$, $\alpha=0$;
(c) coupling function (i) and parameters $\lambda=30$, $\beta=1$, $m_A=0.1826$, $\alpha=0$;
(d) coupling function (i) and parameters $\lambda=40$, $\beta=1$, $m_A=0.158$, 
$\alpha=0$ (solid) and  $\alpha=10$ (dashed). Note the Schwarzschild metric
functions (thin-dotted) for comparison.
}
\label{fig_sol}
\end{figure}

We exhibit some typical vectorized black hole solutions in Fig.~\ref{fig_sol}.
The figures show the metric functions $F_0$ and $f_1$ together with the vector field function $A_t$
versus the compactified radial coordinate $x$, Eq.~(\ref{x_com}), for the coupling functions
(i) and (ii) and selected values of the coupling constants and the potential parameters.
The vector field function exhibits a pronounced maximum at 
several times the horizon radius. This maximum decreases in size 
and shifts to smaller radii as
the vector field mass and self-interaction are increased.
In all cases, the metric functions of the vectorized black holes
deviate only somewhat from the Schwarzschild metric functions,
with the deviation decreasing as the 
vector field mass and self-interaction are increased.

\begin{figure}[h!]
\begin{center}
(a)\includegraphics[width=.45\textwidth, angle =0]{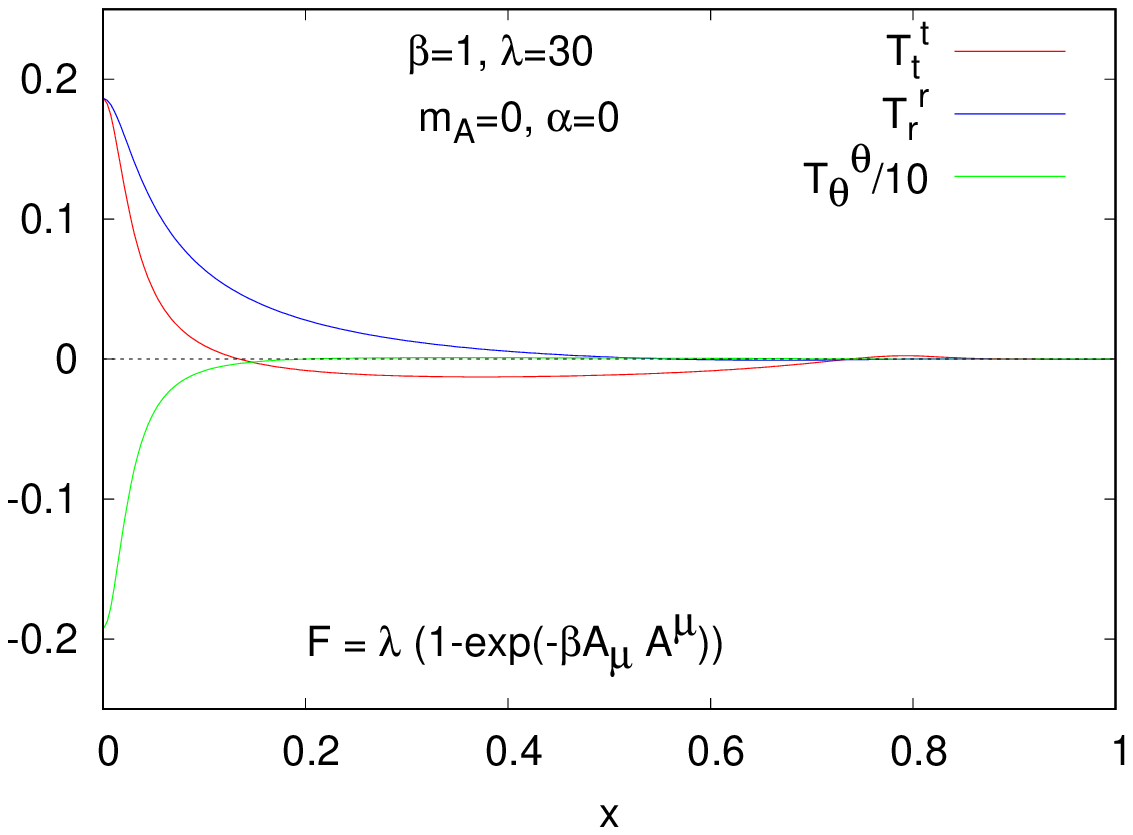}
(b)\includegraphics[width=.45\textwidth, angle =0]{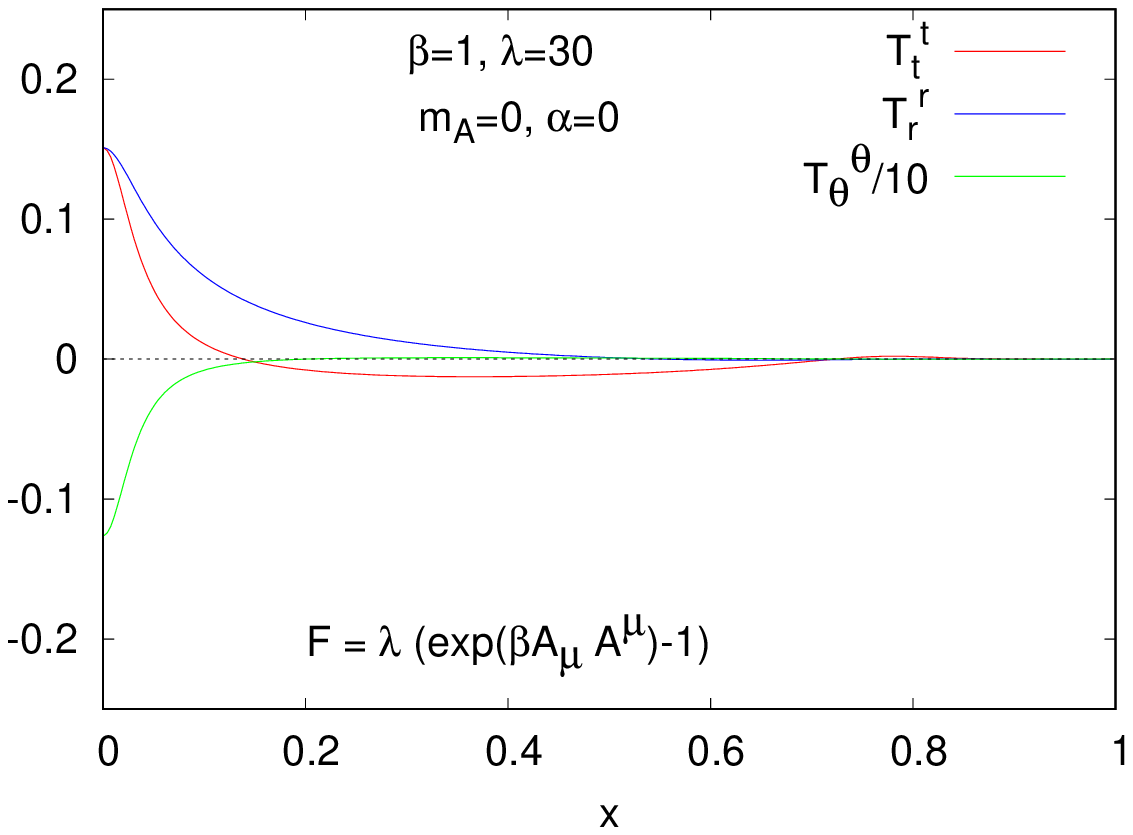}
\\
(c)\includegraphics[width=.45\textwidth, angle =0]{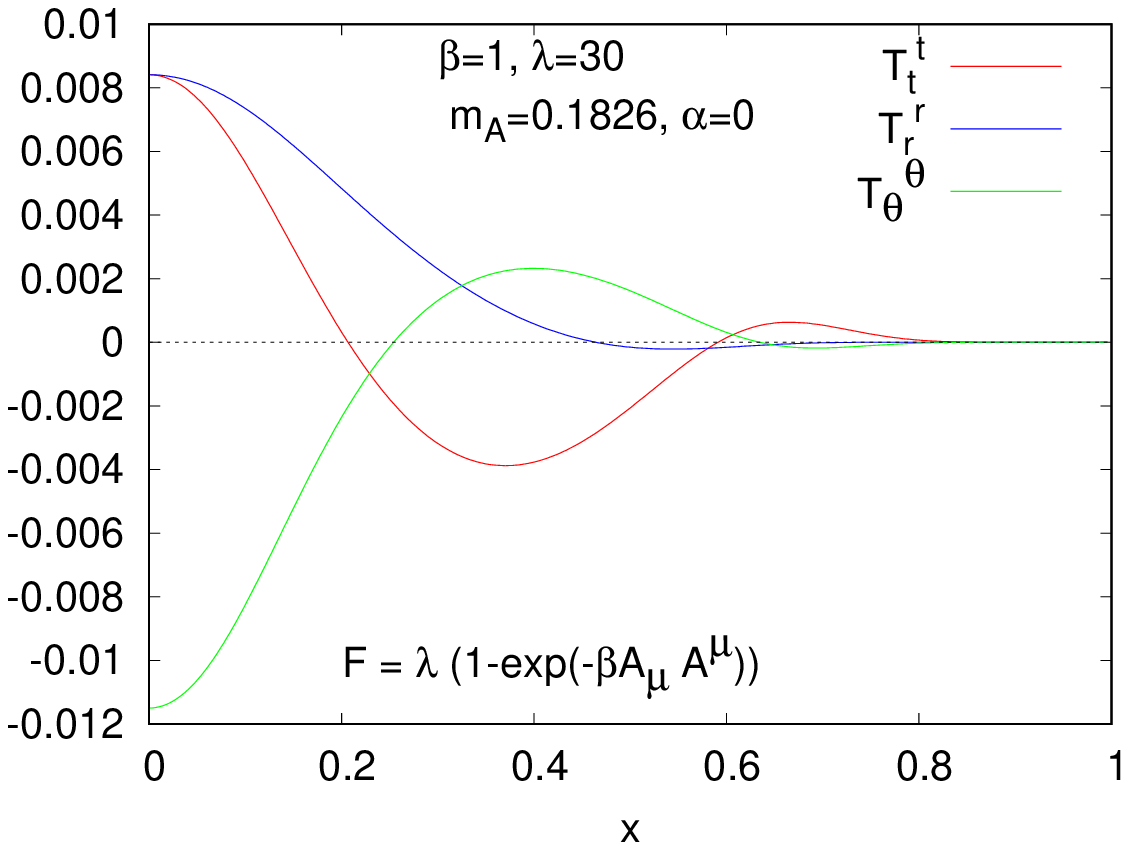}
(d)\includegraphics[width=.45\textwidth, angle =0]{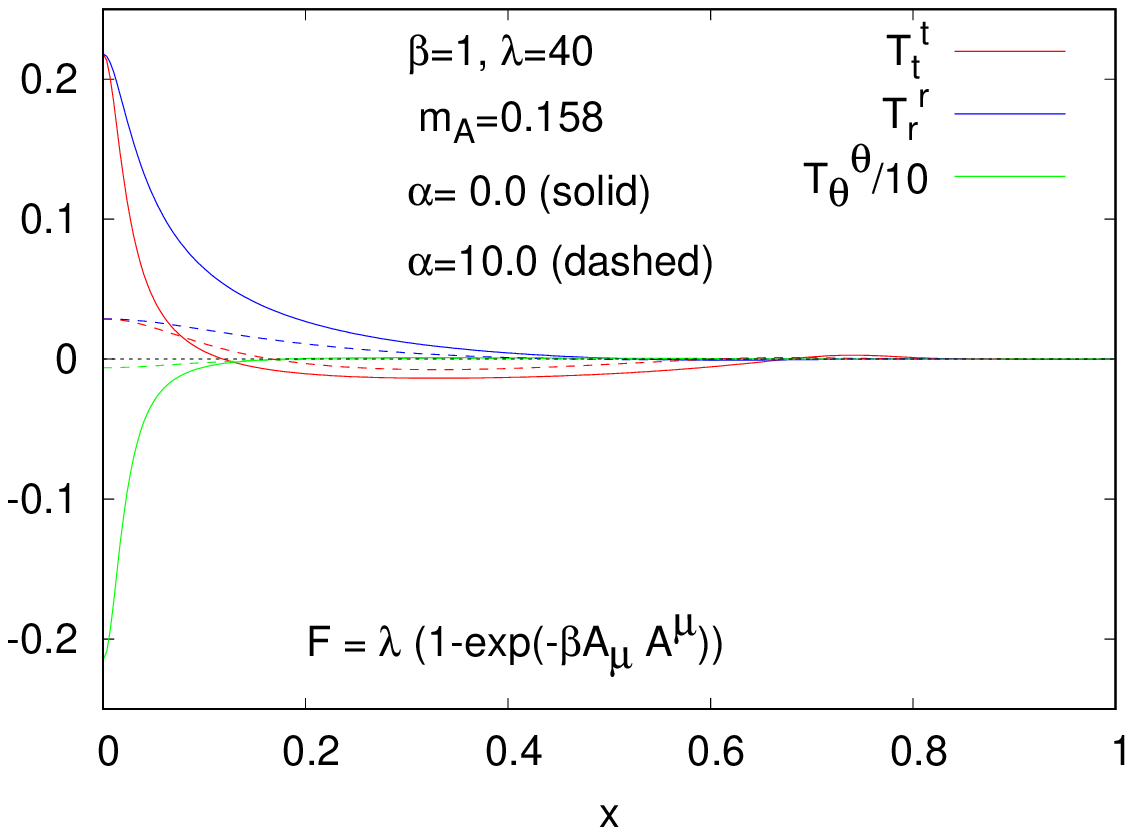}
\end{center}
\caption{Examples of vectorized black hole solutions:
effective stress-energy tensor components $T^{t(\rm eff)}_t$, $T^{r(\rm eff)}_r$, 
and $T^{\theta(\rm eff)}_\theta$
vs compactified radial coordinate $x$ for
(a) coupling function (i) and parameters $\lambda=30$, $\beta=1$, $m_A=0$, $\alpha=0$;
(b) coupling function (ii) and parameters $\lambda=30$, $\beta=1$, $m_A=0$, $\alpha=0$;
(c) coupling function (i) and parameters $\lambda=30$, $\beta=1$, $m_A=0.1826$, $\alpha=0$;
(d) coupling function (i) and parameters $\lambda=40$, $\beta=1$, $m_A=0.158$, 
$\alpha=0$ (solid) and  $\alpha=10$ (dashed).
}
\label{fig_sol2}
\end{figure}

We illustrate for the same set of solutions
the components $T^{t(\rm eff)}_t$, $T^{r(\rm eff)}_r$, 
and $T^{\theta(\rm eff)}_\theta$
of the effective stress-energy tensor
versus the compactified radial coordinate $x$
in Fig.~\ref{fig_sol2}.
We note, that at the horizon 
\begin{eqnarray}
 T^{t(\rm eff)}_t(r_{\rm H}) =  T^{r(\rm eff)}_r(r_{\rm H}) & = &
   4\frac{b_{\rm H}^2}{\rho_{\rm H}^2 f^2_{0 {\rm H}}} 
   \left( 2 l_{\rm H}-1\right) \ ,
\label{t00H} \\
T^{\theta(\rm eff)}_\theta(r_{\rm H}) & = & -T^t_t(r_{\rm H}) 
\frac{f_{0 {\rm H}} + 8 l_{\rm H} b_{\rm H}^2}{f_{0 {\rm H}}} \ ,
\label{tthetathetaH}    
\end{eqnarray}
where we have introduced the circumferential horizon radius 
$\rho_{\rm H} = e^{f_{1,\rm H}/2} r_{\rm H}$, 
and $f_{0 {\rm H}}= f_0(r_{\rm H})$, $b_{\rm H}=b(r_{\rm H})$,
and $l_{\rm H}=\beta\lambda/\rho_{\rm H}^2$
for (i) and (ii), while $l_{\rm H}=\lambda/\rho_{\rm H}^2$ for (iii).
Since $-T^{t(\rm eff)}_t$ can be interpreted
as an effective energy density, Eq.~(\ref{t00H}) shows that near the horizon
the effective energy density is negative. Somewhat away
from the horizon the effective energy density
then turns positive, only to become negative again
when the vector field function approaches its maximum.
Although the effective energy density exhibits this oscillating behaviour,
the contribution to the mass from the region outside the horizon
is in all cases positive.

\subsubsection{Domain of existence: massless case}

\begin{figure}[h!]
\begin{center}
\includegraphics[width=.45\textwidth, angle =0]{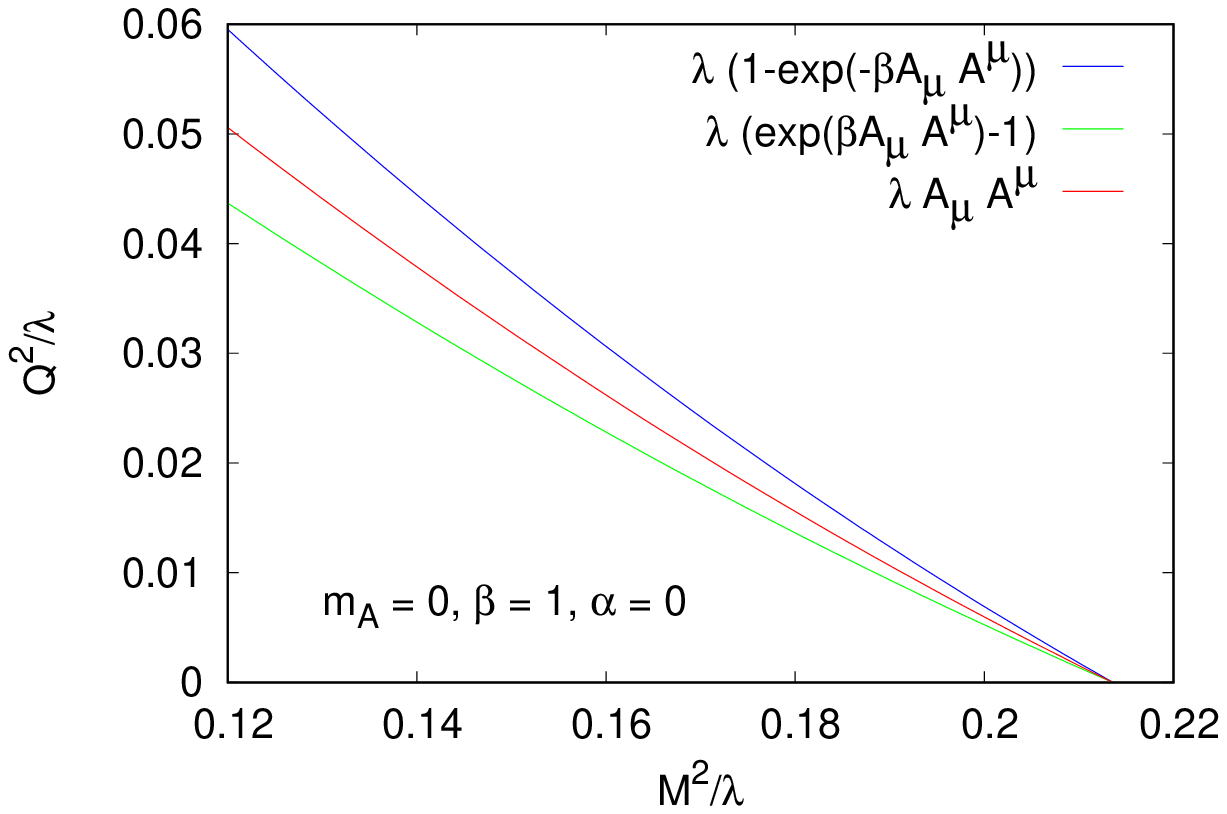}
\includegraphics[width=.45\textwidth, angle =0]{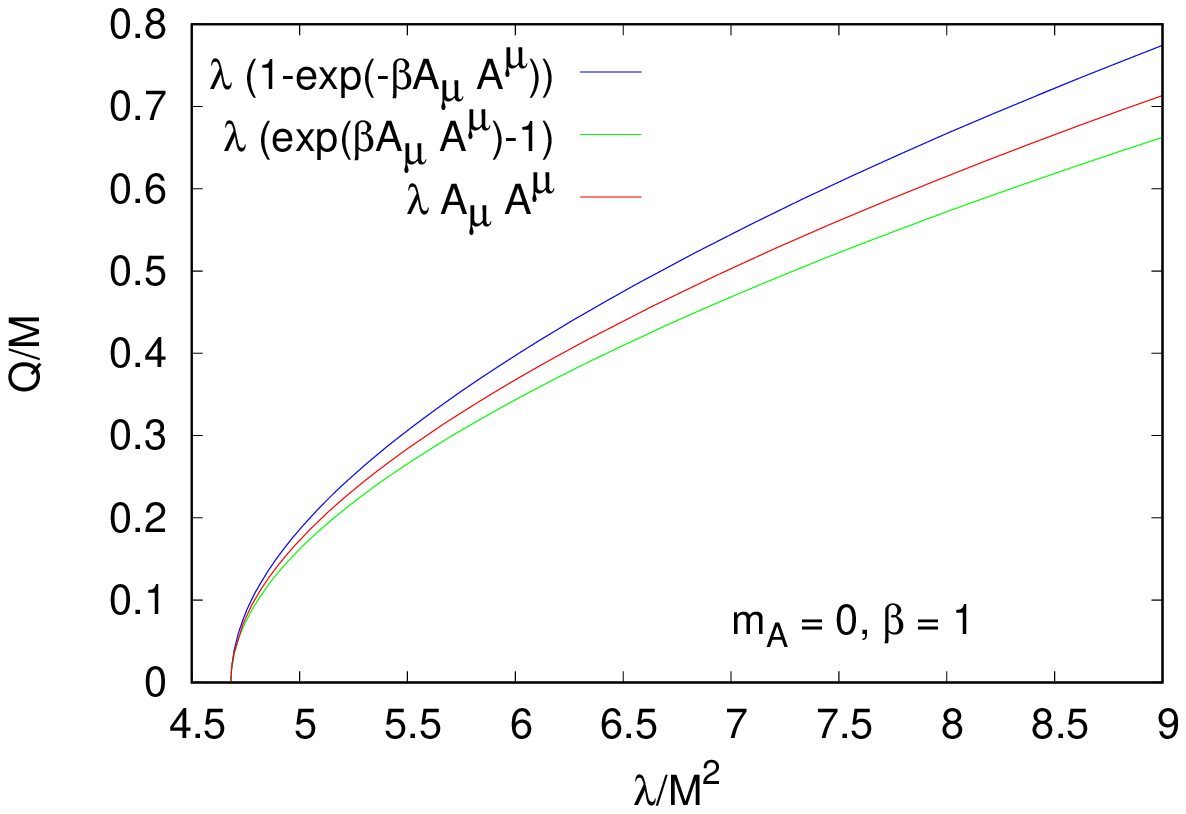}
\end{center}
\caption{Domain of existence for all three coupling functions ($V=0$):
(a) vector charge $Q^2/\lambda$ vs black hole mass $M^2/\lambda$;
(b) vector charge $Q/M$ vs coupling constant $\lambda/M^2$.
}
\label{fig_dom1}
\end{figure}

We now address the domain of existence of the vectorized black holes
for vanishing potential $V$.
The domain of existence is illustrated in Fig.~\ref{fig_dom1}
for all three coupling functions, where we have set the second
coupling constant $\beta$ for the cases (i) and (ii) to $\beta=1$.
We show in Fig.~\ref{fig_dom1}(a) the
vector charge $Q^2/\lambda$ versus the black hole mass $M^2/\lambda$,
where we have scaled with the coupling constant $\lambda$.
For comparison we show
in Fig.~\ref{fig_dom1}(b) the vector charge $Q/M$ versus the coupling constant $\lambda/M^2$,
where we have scaled with the black hole mass $M$.

We note, that independent of the coupling function,
the branches of vectorized black holes emerge from the Schwarzschild solution
at $M^2/\lambda=0.2136$, where the tachyonic instability of the Schwarzschild
solution sets in, manifesting in a zero mode of the Schwarzschild solution.
The branches then extend to smaller values of $M^2/\lambda$.
Here the effect of the coupling function becomes important, and we note,
that the vector charge $Q/M$ is largest for the coupling function (i), 
and smallest for the coupling function (ii).
The branches finally end at critical solutions,  
when $M^2/\lambda$ tends to zero.
At these critical solutions a curvature singularity is encountered at the horizon.

To analyze the critical behaviour we consider the Ricci scalar  and the GB invariant
at the horizon, $R_{\rm H}$ and $R^2_{\rm GB \, H}$, respectively, and scale these
with the square of the circumferential horizon radius 
$\rho_{\rm H}$, 
to obtain scale-invariant expressions.
Analytic expressions for these scaled curvature invariants at the horizon are then given by
\begin{eqnarray}
\rho_{\rm H}^2 R_{\rm H} & = & 
\frac{64 b_{\rm H}^4}{f_{0 {\rm H}}^2}
l_{\rm H}
\left(2 l_{\rm H} - 1\right) ,
\nonumber\\
\rho_{\rm H}^4 R^2_{\rm GB \, H} & = & 
\frac{4}{f_{0 {\rm H}}^2}
\left\{
 16  b_{\rm H}^4 \left(2 l_{\rm H} - 1\right)
     \left(6 l_{\rm H}  - 1\right)
      +3 \left[8 \left(2 l_{\rm H}  - 1 \right) b_{\rm H}^2 
      + f_{0 {\rm H}} \right] f_{0 {\rm H}} 
 \right\} ,
\label{sccurvi}
\end{eqnarray}
in the notation of Eqs.~(\ref{t00H})-(\ref{tthetathetaH}).
Note that these expressions are independent of the potential $V(A_\mu A^\mu)$.

\begin{figure}[h!]
\begin{center}
\includegraphics[width=.45\textwidth, angle =0]{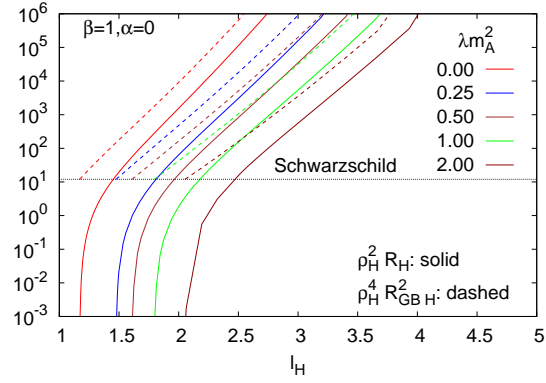}
\end{center}
\caption{Curvature invariants at the horizon:
scale-invariant curvature invariants $\rho_{\rm H}^2 R_{\rm H}$ (solid) and 
$\rho_{\rm H}^4 R^2_{\rm GB \, H}$ (dashed) vs the dimensionless
coupling parameter $l_{\rm H}$ for coupling function $(i)$ and several values of the 
vector field mass $m_{\rm A}$.
}
\label{singH}
\end{figure}

We demonstrate the critical behaviour in Fig.~\ref{singH} for the coupling function (i),
where we show the scaled Ricci scalar  
$\rho_{\rm H}^2 R_{\rm H}$ and the scaled GB invariant $\rho_{\rm H}^2 R^2_{\rm GB \, H}$
at the horizon as functions of the scaled coupling parameter 
$l_{\rm H}=\beta\lambda/\rho_{\rm H}^2$.
We observe that these scaled  
curvature invariants increase exponentially with $l_{\rm H}$ and reach
very large values already for moderate values of $l_{\rm H}$.

As in the case of scalarization, there are also excited vectorized black hole solutions.
Here we only note that independent of the coupling function
the branches of vectorized black holes with a single node arise
at $M^2/\lambda=0.00598$.
This is to be compared to the onset of the fundamental branches of vectorized solutions
at $M^2/\lambda=0.2136$.
We expect a countable number of higher excited solutions, arising at successively smaller
values of $M^2/\lambda$.

\begin{figure}[h!]
\begin{center}
\includegraphics[width=.45\textwidth, angle =0]{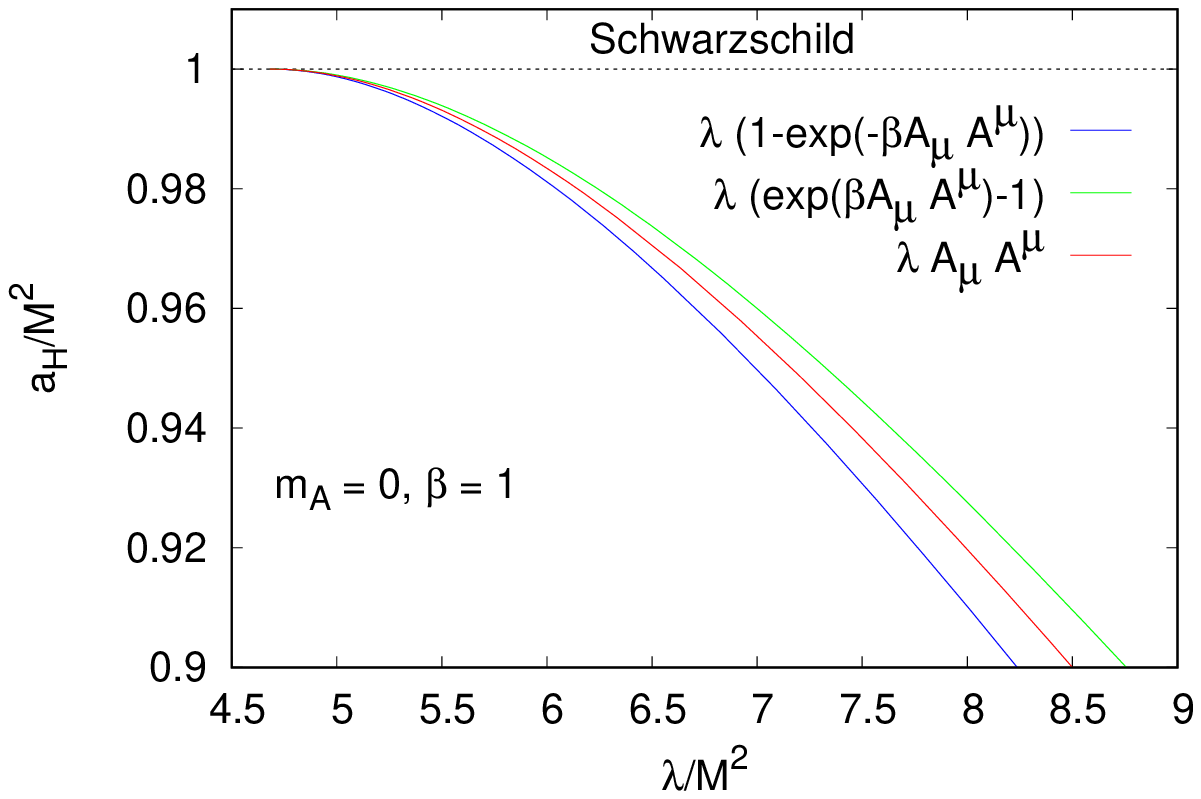}
\includegraphics[width=.45\textwidth, angle =0]{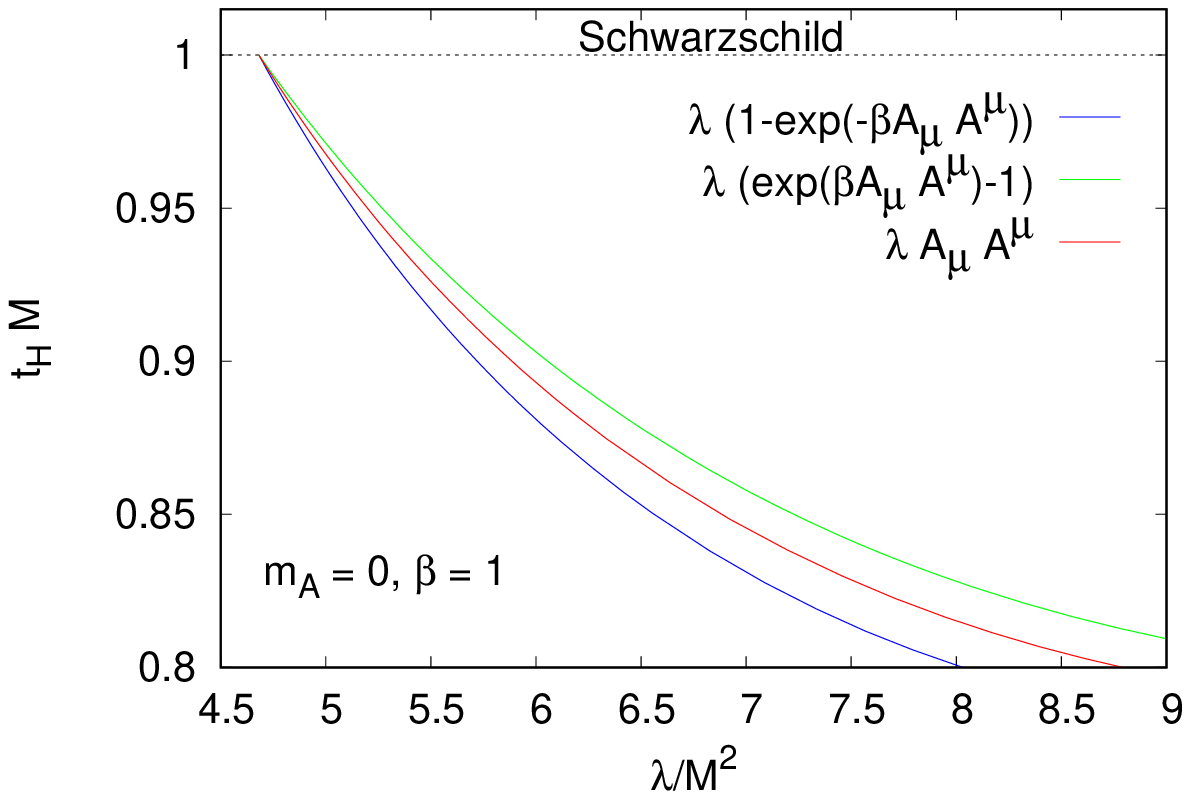}
\end{center}
\caption{Horizon properties for all three coupling functions ($V=0$):
(a) reduced horizon area $a_{\rm H}/M^2$ vs coupling constant $\lambda/M^2$;
(b) reduced temperature $t_{\rm H} M$ vs coupling constant $\lambda/M^2$.
}
\label{fig_hor1}
\end{figure}

We now turn to the horizon properties and 
define the reduced horizon area  $a_{\rm H}$ and the reduced temperature $t_{\rm H}$
\begin{equation}
a_{\rm H} = \frac{A_{\rm H}}{16 \pi} , \ \ \ \
t_{\rm H} = 4 \pi T_{\rm H} .
\label{reduced}
\end{equation}
We exhibit in Fig.~\ref{fig_hor1}(a) the reduced horizon area $a_{\rm H}/M^2$ 
and in Fig.~\ref{fig_hor1}(b ) the  reduced temperature $t_{\rm H} M$ 
for all three coupling functions and vanishing potential $V$ 
versus the  coupling constant $\lambda/M^2$.
For all three coupling functions the area of the vectorized black holes
is smaller than for the Schwarzschild black holes,
and the area is smallest for the coupling function (i) and largest for (ii).
Analogously the temperature of the vectorized black holes
is smaller than for the Schwarzschild black holes.

As discussed above, the entropy of these vectorized black holes
is simply given by a quarter of their horizon area,
since the GB term does not contribute,
being multiplied by a coupling function
which vanishes at the horizon.
Thus we have to conclude from Fig.~\ref{fig_hor1}(a)
that the Schwarzschild solutions are entropically favored over 
the vectorized solutions.
The reason, that the horizon area and thus the entropy is smaller
for the vectorized solutions then stems from the fact that
their total mass contains a contribution from the bulk
outside the horizon. Therefore, for a given total mass,
the horizon radius for a vectorized black hole
is smaller than for a Schwarzschild back hole,
which is a vacuum solution.

\subsubsection{Domain of existence: influence of vector field mass $m_A$} 

\begin{figure}[h!]
\begin{center}
\includegraphics[width=.45\textwidth, angle =0]{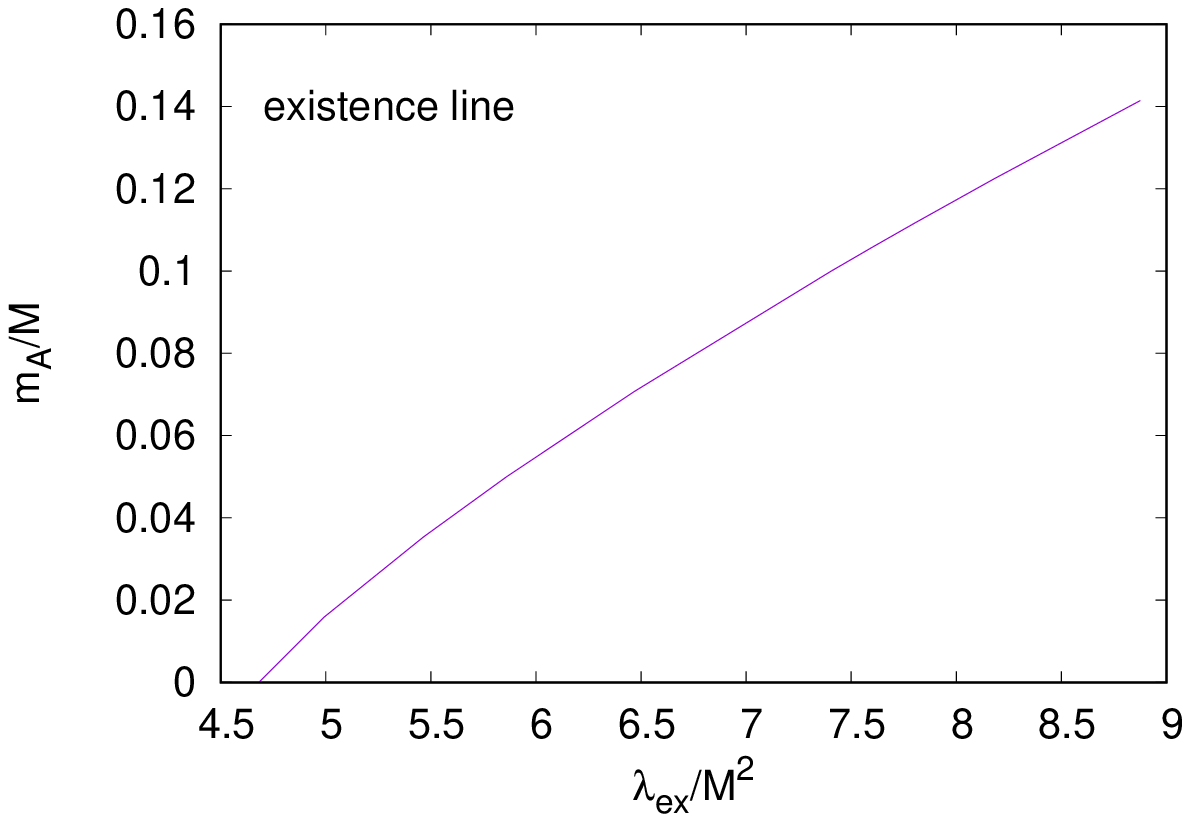}
\includegraphics[width=.45\textwidth, angle =0]{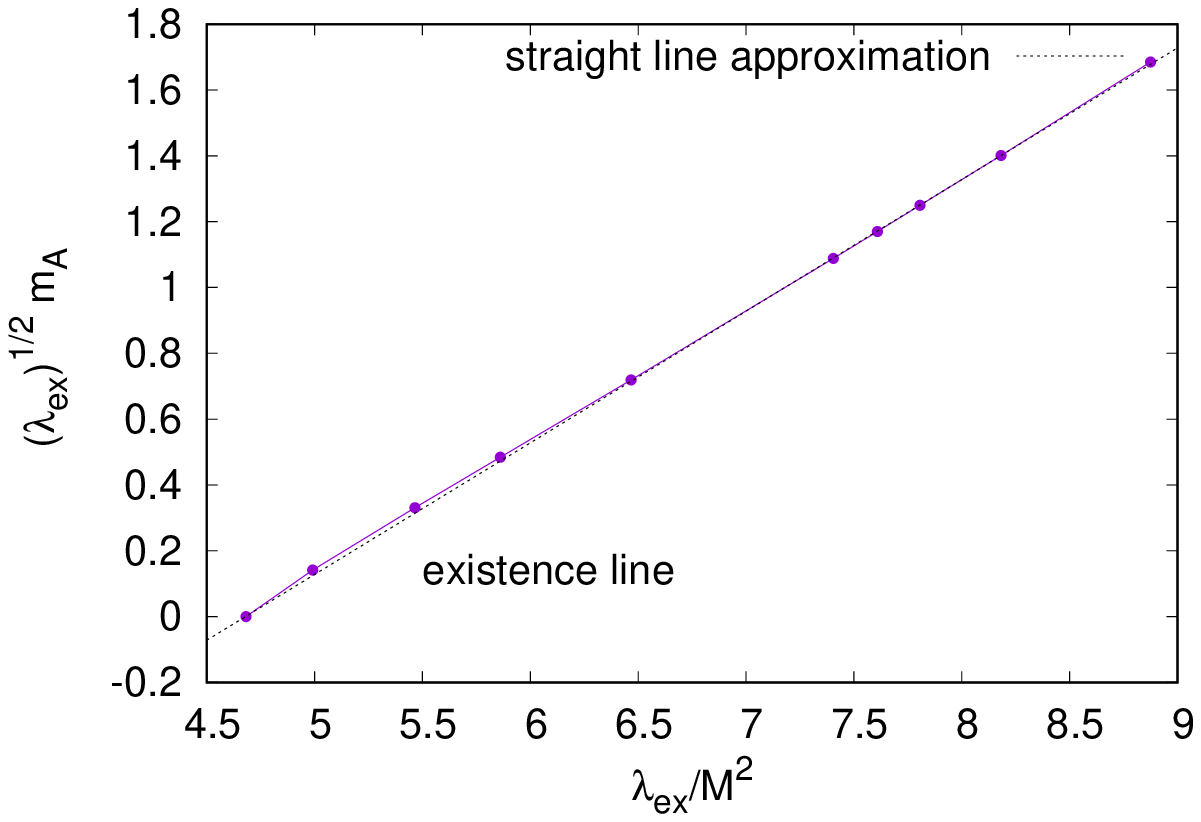}
\end{center}
\caption{Existence line:
(a) vector field mass $m_A/M$ vs coupling constant $\lambda_{\rm ex}/M^2$;
(b) vector field mass $\lambda_{\rm ex}^{1/2} m_A$ vs coupling constant $\lambda_{\rm ex}/M^2$
(solid) together with a linear fit (dotted).
}
\label{fig_ex}
\end{figure}

We next consider the effects of a finite mass $m_A$ of the vector field.
The presence of an ordinary finite mass term in the vector field equation (\ref{scleq})
clearly affects the effective mass responsible for the tachyonic instability
of the Schwarzschild black holes, which now consists of two contributions:
the ordinary mass and the curvature-induced mass.
Consequently, the value of the coupling constant $\lambda$,
where the Schwarzschild solution develops a zero mode, 
changes with the vector field mass $m_A$.
Denoting this coupling constant by $\lambda_{\rm ex}$,
we thus obtain the existence line for the vectorized black hole solutions $m_A(\lambda_{\rm ex})$.

We exhibit the existence line in Fig.~\ref{fig_ex}.
We show the vector field mass $m_A/M$ versus the coupling constant $\lambda_{\rm ex}/M^2$
 in Fig.~\ref{fig_ex}(a).
The figure shows, that the onset of the tachyonic instability of the Schwarzschild black hole 
is shifted to larger values of $\lambda$, when the vector field mass is increased.
This is to be expected, since the finite vector field mass increases the effective mass in
the vector field equation, which must then be compensated by a larger contribution
from the curvature-induced contribution to the effective mass, 
and this latter contribution is proportional to $\lambda$.
When considering the vector field mass $\lambda_{\rm ex}^{1/2} m_A$ versus the
coupling constant $\lambda_{\rm ex}/M^2$, as shown in Fig.~\ref{fig_ex}(b),
we obtain basically a linear relation, also demonstrated in the figure by the linear fit.

The existence line depends only on the effective mass in the vector field equation (\ref{scleq})
and thus the terms linear in the vector field.
Higher powers of the vector field do not matter for the onset of the tachyonic instability.
Therefore all three coupling functions possess the same existence line.
Similarly, adding self-interaction terms to the potential $V$ will also not affect
the existence line.
The effect of higher powers in the coupling function or in the potential
does of course influence the domain of existence of the vectorized black hole solutions.

\begin{figure}[h!]
\begin{center}
\includegraphics[width=.45\textwidth, angle =0]{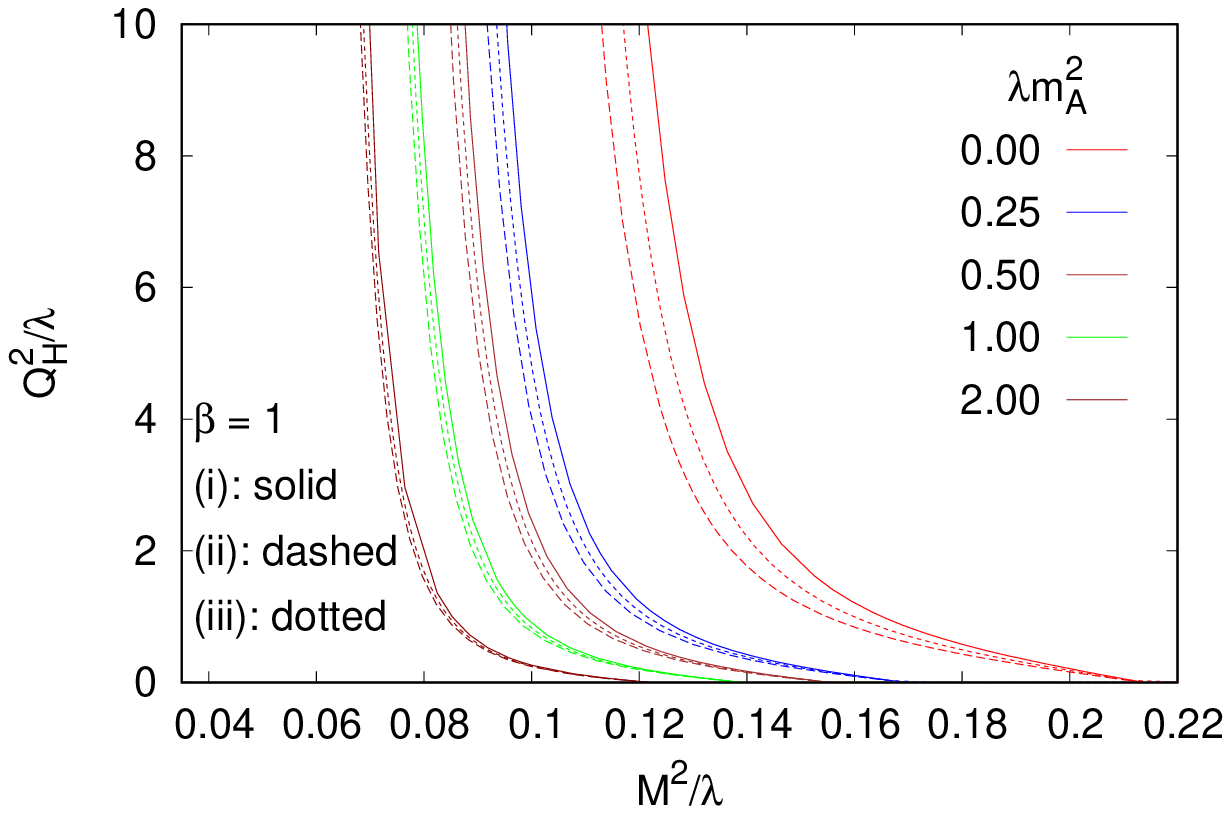}
\includegraphics[width=.45\textwidth, angle =0]{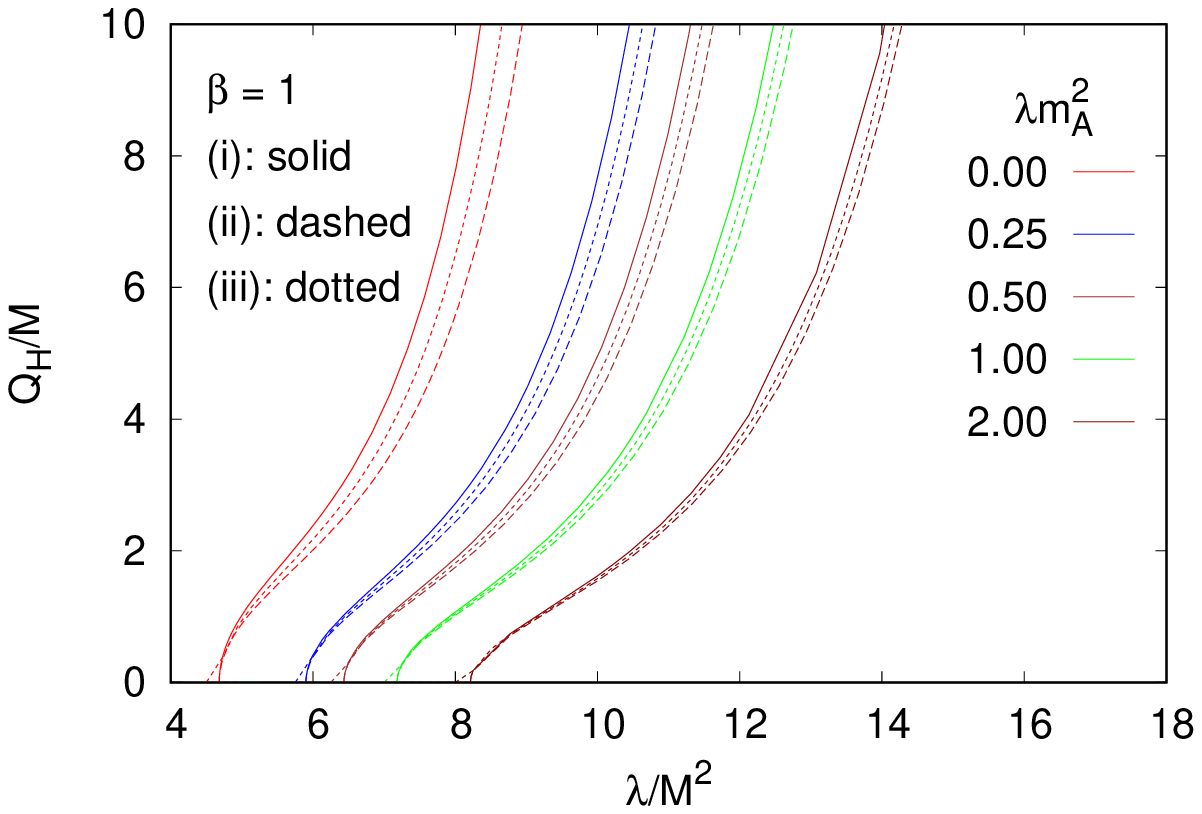}
\end{center}
\caption{Domain of existence for coupling functions (i), (ii) and (iii) ($V = 2 m_A^2 A_\mu A^\mu$)
for several values of $\lambda m_A^2$:
(a) vector charge $Q_{\rm H}^2/\lambda$ vs black hole mass $M^2/\lambda$;
(b) vector charge $Q_{\rm H}/M$ vs coupling constant $\lambda/M^2$.
}
\label{fig_dom2}
\end{figure}

We exhibit in Fig.~\ref{fig_dom2} the domain of existence of the vectorized black hole
solutions for the  coupling functions (i), (ii) and (iii) with potential
$V = 2 m_A^2 A_\mu A^\mu$, to illustrate
the dependence on the vector field mass $m_A$.
Since the charge $Q$ vanishes
for solutions with non-zero vector field mass, we employ
the horizon charge $Q_{\rm H}$, Eq.~(\ref{horzcharge}).
Fig.~\ref{fig_dom2}(a) shows
the horizon charge $Q_{\rm H}^2/\lambda$ versus the black hole mass $M^2/\lambda$, 
and Fig.~\ref{fig_dom2}(b) the
horizon charge $Q_{\rm H}/M$ versus the coupling constant $\lambda/M^2$.
Analogously to the case of vanishing vector field mass, the vectorized black hole solutions
develop a curvature singularity at the horizon when $M^2/\lambda$ tends to zero.
As noted above, the scaled curvature invariants at the horizon, Eqs.~(\ref{sccurvi}),
are independent of the potential $V(A_\mu A^\mu)$. 
The dependence
of the scaled curvature invariants on the vector field mass $m_A$ is seen in
Fig.~\ref{singH}. 

\begin{figure}[h!]
\begin{center}
\includegraphics[width=.45\textwidth, angle =0]{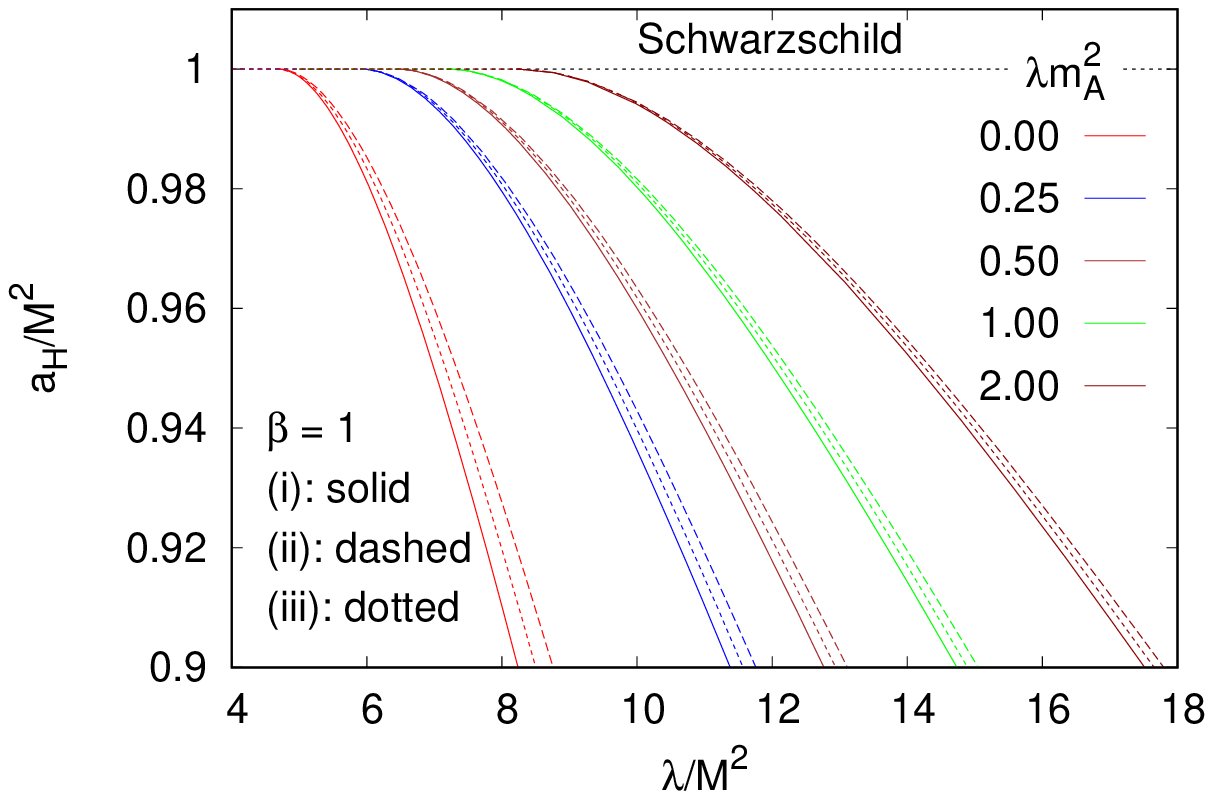}
\includegraphics[width=.45\textwidth, angle =0]{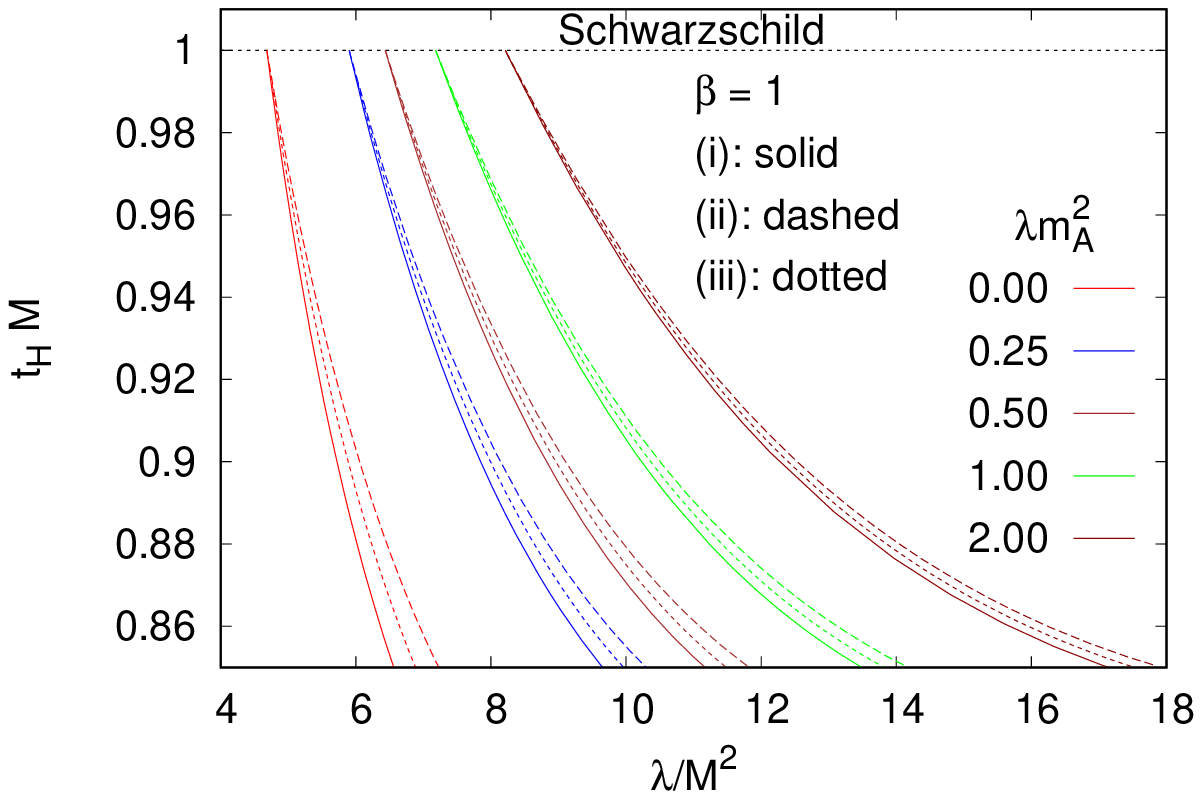}
\end{center}
\caption{Horizon properties for coupling functions (i), (ii) and (iii) ($V = 2 m_A^2 A_\mu A^\mu$)
for several values of $\lambda m_A^2$:
(a) reduced horizon area $a_{\rm H}/M^2$ vs coupling constant $\lambda/M^2$;
(b) reduced temperature $t_{\rm H} M$ vs coupling constant $\lambda/M^2$.
}
\label{fig_hor2}
\end{figure}

We illustrate the dependence of the horizon properties 
on the vector field mass $m_A$ in Fig.~\ref{fig_hor2}
for the  coupling functions (i), (ii) and (iii) with potential
$V = 2 m_A^2 A_\mu A^\mu$.
The reduced horizon area $a_{\rm H}/M^2$ is shown in Fig.~\ref{fig_hor2}(a) 
and the reduced temperature $t_{\rm H} M$ in Fig.~\ref{fig_hor2}(b).  
We note that also in the presence of a finite vector field mass $m_A$
the horizon area of the vectorized black holes
is smaller than for the Schwarzschild black holes,
and the area is smallest for the coupling function (i) and largest for (ii).
But the area $a_{\rm H}/M^2$ decreases less rapidly with increasing $\lambda/M^2$,
when the vector field mass increases.
The temperature exhibits a similar dependence on the vector field mass.
 
Since the entropy of these vectorized black holes 
is simply given by a quarter of their horizon area,
independent of the potential $V$,
we conclude from Fig.~\ref{fig_hor2}(a)
that as in the massless case
the Schwarzschild black holes are entropically favored
over the vectorized black holes,
independent of the employed coupling function.

\section{Conclusions}

Here we have performed a first exploratory study of 
curvature-induced spontaneously vectorized black holes.
These novel black holes arise when a vector field
is coupled to the GB term
by employing a coupling function that is quadratic
in the vector field.
The GB term then induces a tachyonic instability
of the Schwarzschild black holes, which start to grow vector hair.

We have allowed for three different types of coupling functions
in order to see their basic influence.
However, unlike the case of 
curvature-induced spontaneously scalarized black holes,
we did not observe distinctly different physical properties
of the vectorized black holes
for these coupling functions.
In particular, for all coupling functions
the branches of vectorized black holes
extend from their bifurcation point
to smaller values of the scaled coupling constant $M^2/\lambda$.

The bifurcation point does not depend on the coupling function.
However it does depend on the mass of the vector field.
With increasing mass the bifurcation point shifts to larger values
of the coupling constant, while it is not affected by vector 
field self-interactions. 
Radially excited spontaneously vectorized black holes also exist.
Their bifurcation points are at smaller values of $M^2/\lambda$
than the bifurcation point of the fundamental branch of vectorized black holes.

We have shown that
independent of the coupling function and the vector field potential $V$,
all fundamental branches extend to $M^2/\lambda \to 0$,
where the vectorized black holes develop a curvature singularity
at the horizon. The scaled Ricci scalar and the scaled GB invariant
exhibit an exponential dependence on the scaled coupling parameter,
which results in a divergence in the limit.

We have also addressed some thermodynamic properties of 
the vectorized black holes.
In particular, we have calculated the horizon temperature,
the horizon area and the entropy.
Along the branches of vectorized black holes
the scaled temperature and the scaled entropy
decrease monotonically with increasing coupling constant
from the Schwarzschild values at the bifurcation.
Thus for a given mass, the vectorized black holes
have smaller area than the Schwarzschild black holes.
Since the GB term of the vectorized black holes
does not contribute to their entropy,
this entails that Schwarzschild black holes
are entropically preferred.

There are various interesting directions to continue these
investigations. These include foremost a mode analysis of
the static spherically symmetric vectorized black holes
and a generalization to the rotating case.
But also a further analysis of the physical properties
is called for, ranging from a study of their geodesics
and lightrings 
to their accretion discs.

\section*{Acknowledgement}

JK is grateful for valuable discussions with
Kamal Hajian on the BH entropy in the presence of a GB term.
BH, BK and JK gratefully acknowledge support by the
DFG Research Training Group 1620 {\sl Models of Gravity}
and the COST Action CA16104. 
BH acknowledges support from FAPESP under grant number 2019/01511-5.


\begin{thebibliography}{unsrtnat}

\bibitem{Chrusciel:2012jk}
P.~T.~Chrusciel, J.~Lopes Costa and M.~Heusler,
Living Rev. Rel. \textbf{15}, 7 (2012)

\bibitem{Herdeiro:2015waa}
C.~A.~R.~Herdeiro and E.~Radu,
Int. J. Mod. Phys. D \textbf{24}, no.09, 1542014 (2015)

\bibitem{Zwiebach:1985uq}
  B.~Zwiebach,
  Phys.\ Lett.\  {\bf 156B}, 315 (1985).

\bibitem{Gross:1986mw}
  D.~J.~Gross and J.~H.~Sloan,
  Nucl.\ Phys.\ B {\bf 291}, 41 (1987).

\bibitem{Metsaev:1987zx}
  R.~R.~Metsaev, A.~A.~Tseytlin,
  Nucl.\ Phys.\  {\bf B293 }, 385 (1987).

\bibitem{Horndeski:1974wa} 
  G.~W.~Horndeski,
  Int.\ J.\ Theor.\ Phys.\  {\bf 10}, 363 (1974).

\bibitem{Charmousis:2011bf} 
  C.~Charmousis, E.~J.~Copeland, A.~Padilla and P.~M.~Saffin,
  Phys.\ Rev.\ Lett.\  {\bf 108}, 051101 (2012)

\bibitem{Kobayashi:2011nu} 
  T.~Kobayashi, M.~Yamaguchi and J.~Yokoyama,
  Prog.\ Theor.\ Phys.\  {\bf 126}, 511 (2011)
  
\bibitem{Kanti:1995vq}
  P.~Kanti, N.~E.~Mavromatos, J.~Rizos, K.~Tamvakis and E.~Winstanley,
  Phys.\ Rev.\  D {\bf 54} (1996) 5049.

\bibitem{Torii:1996yi}
T.~Torii, H.~Yajima and K.~i.~Maeda,
Phys.\ Rev.\ D {\bf 55}, 739 (1997)

\bibitem{Guo:2008hf}
  Z.~K.~Guo, N.~Ohta and T.~Torii,
  Prog.\ Theor.\ Phys.\  {\bf 120}, 581 (2008)

\bibitem{Pani:2009wy}
P.~Pani and V.~Cardoso,
Phys.\ Rev.\ D {\bf 79}, 084031 (2009)

\bibitem{Pani:2011gy}
  P.~Pani, C.~F.~B.~Macedo, L.~C.~B.~Crispino and V.~Cardoso,
  Phys.\ Rev.\ D {\bf 84}, 087501 (2011)

\bibitem{Kleihaus:2011tg}
  B.~Kleihaus, J.~Kunz and E.~Radu,
  Phys.\ Rev.\ Lett.\  {\bf 106} (2011) 151104.

\bibitem{Ayzenberg:2013wua}
  D.~Ayzenberg, K.~Yagi and N.~Yunes,
  Phys.\ Rev.\ D {\bf 89}, no. 4, 044023 (2014)

\bibitem{Ayzenberg:2014aka}
D.~Ayzenberg and N.~Yunes,
Phys.\ Rev.\ D {\bf 90}, 044066 (2014)

\bibitem{Maselli:2015tta}
A.~Maselli, P.~Pani, L.~Gualtieri and V.~Ferrari,
Phys.\ Rev.\ D {\bf 92}, no. 8, 083014 (2015)

\bibitem{Kleihaus:2014lba}
  B.~Kleihaus, J.~Kunz and S.~Mojica,
  Phys.\ Rev.\ D {\bf 90}, no. 6, 061501 (2014)

\bibitem{Kleihaus:2015aje}
  B.~Kleihaus, J.~Kunz, S.~Mojica and E.~Radu,
  Phys.\ Rev.\ D {\bf 93}, no. 4, 044047 (2016)

\bibitem{Blazquez-Salcedo:2016enn} 
  J.~L.~Bl\'{a}zquez-Salcedo, C.~F.~B.~Macedo, V.~Cardoso, V.~Ferrari, L.~Gualtieri, F.~S.~Khoo, J.~Kunz and P.~Pani,
  Phys.\ Rev.\ D {\bf 94}, no. 10, 104024 (2016)
  
\bibitem{Blazquez-Salcedo:2017txk}
J.~L.~Bl\'azquez-Salcedo, F.~S.~Khoo and J.~Kunz,
Phys. Rev. D \textbf{96}, no.6, 064008 (2017)

\bibitem{Sotiriou:2013qea} 
  T.~P.~Sotiriou and S.~Y.~Zhou,
  Phys.\ Rev.\ Lett.\  {\bf 112}, 251102 (2014)

\bibitem{Sotiriou:2014pfa} 
  T.~P.~Sotiriou and S.~Y.~Zhou,
  Phys.\ Rev.\ D {\bf 90}, 124063 (2014)
  
\bibitem{Antoniou:2017acq}
G.~Antoniou, A.~Bakopoulos and P.~Kanti,
Phys.\ Rev.\ Lett.\  {\bf 120}, no. 13, 131102 (2018);
Phys.\ Rev.\ D {\bf 97} (2018) no.8,  084037.

\bibitem{Doneva:2017bvd}
D.~D.~Doneva and S.~S.~Yazadjiev,
Phys.\ Rev.\ Lett.\  {\bf 120}, no. 13, 131103 (2018).

\bibitem{Silva:2017uqg}
H.~O.~Silva, J.~Sakstein, L.~Gualtieri, T.~P.~Sotiriou and E.~Berti,
Phys.\ Rev.\ Lett.\  {\bf 120}, no. 13, 131104 (2018).

\bibitem{Antoniou:2017hxj} 
  G.~Antoniou, A.~Bakopoulos and P.~Kanti,
  Phys.\ Rev.\ D {\bf 97}, no. 8, 084037 (2018)

\bibitem{Blazquez-Salcedo:2018jnn} 
  J.~L.~Bl\'{a}zquez-Salcedo, D.~D.~Doneva, J.~Kunz and S.~S.~Yazadjiev,
  Phys.\ Rev.\ D {\bf 98}, no. 8, 084011 (2018).

\bibitem{Doneva:2018rou}
D.~D.~Doneva, S.~Kiorpelidi, P.~G.~Nedkova, E.~Papantonopoulos and S.~S.~Yazadjiev,
Phys.\ Rev.\ D \textbf{98}, no.10, 104056 (2018)

\bibitem{Minamitsuji:2018xde} 
  M.~Minamitsuji and T.~Ikeda,
  Phys.\ Rev.\ D {\bf 99}, no. 4, 044017 (2019)

\bibitem{Silva:2018qhn} 
  H.~O.~Silva, C.~F.~B.~Macedo, T.~P.~Sotiriou, L.~Gualtieri, J.~Sakstein and E.~Berti,
  Phys.\ Rev.\ D {\bf 99}, no. 6, 064011 (2019)

\bibitem{Brihaye:2018grv} 
  Y.~Brihaye and L.~Ducobu,
  Phys.\ Lett.\ B {\bf 795}, 135 (2019)

\bibitem{Bakopoulos:2018nui}
 A.~Bakopoulos, G.~Antoniou and P.~Kanti,
  Phys.\ Rev.\ D {\bf 99}, no.6,  064003 (2019)

\bibitem{Doneva:2019vuh} 
  D.~D.~Doneva, K.~V.~Staykov and S.~S.~Yazadjiev,
  Phys.\ Rev.\ D {\bf 99}, no. 10, 104045 (2019)

\bibitem{Macedo:2019sem} 
  C.~F.~B.~Macedo, J.~Sakstein, E.~Berti, L.~Gualtieri, H.~O.~Silva and T.~P.~Sotiriou,
  Phys.\ Rev.\ D {\bf 99}, no. 10, 104041 (2019)

\bibitem{Myung:2019wvb} 
  Y.~S.~Myung and D.~C.~Zou,
  Int.\ J.\ Mod.\ Phys.\ D {\bf 28}, no. 09, 1950114 (2019)

\bibitem{Cunha:2019dwb} 
  P.~V.~P.~Cunha, C.~A.~R.~Herdeiro and E.~Radu,
  Phys.\ Rev.\ Lett.\  {\bf 123}, no. 1, 011101 (2019)

\bibitem{Bakopoulos:2019tvc}
  A.~Bakopoulos, P.~Kanti and N.~Pappas,
  Phys.\ Rev.\ D {\bf 101}, no.4,  044026 (2020)

\bibitem{Hod:2019pmb}
  S.~Hod,
  Phys.\ Rev.\ D {\bf 100} (2019) no.6,  064039

\bibitem{Collodel:2019kkx}
L.~G.~Collodel, B.~Kleihaus, J.~Kunz and E.~Berti,
Class.\ Quant.\ Grav.\  \textbf{37}, no.7, 075018 (2020)

\bibitem{Bakopoulos:2020dfg}
A.~Bakopoulos, P.~Kanti and N.~Pappas,
Phys.\ Rev.\ D {\bf 101}, no.8,  084059 (2020)

\bibitem{Blazquez-Salcedo:2020rhf}
J.~L.~Bl\'azquez-Salcedo, D.~D.~Doneva, S.~Kahlen, J.~Kunz, P.~Nedkova and S.~S.~Yazadjiev,
Phys. Rev. D \textbf{101} (2020) no.10, 104006

\bibitem{Blazquez-Salcedo:2020caw}
J.~L.~Bl\'azquez-Salcedo, D.~D.~Doneva, S.~Kahlen, J.~Kunz, P.~Nedkova and S.~S.~Yazadjiev,
[arXiv:2006.06006 [gr-qc]].

\bibitem{Herdeiro:2020wei}
C.~A.~R.~Herdeiro, E.~Radu, H.~O.~Silva, T.~P.~Sotiriou and N.~Yunes,
[arXiv:2009.03904 [gr-qc]].

\bibitem{Berti:2020kgk}
E.~Berti, L.~G.~Collodel, B.~Kleihaus and J.~Kunz,
[arXiv:2009.03905 [gr-qc]].

\bibitem{Dima:2020yac}
A.~Dima, E.~Barausse, N.~Franchini and T.~P.~Sotiriou,
Phys. Rev. Lett. \textbf{125}, no.23, 231101 (2020)

\bibitem{Hod:2020jjy}
S.~Hod,
Phys. Rev. D \textbf{102}, no.8, 084060 (2020)

\bibitem{Doneva:2020nbb}
D.~D.~Doneva, L.~G.~Collodel, C.~J.~Kr\"uger and S.~S.~Yazadjiev,
Phys. Rev. D \textbf{102}, 104027 (2020)

\bibitem{Doneva:2020kfv}
D.~D.~Doneva, L.~G.~Collodel, C.~J.~Kr\"uger and S.~S.~Yazadjiev,
[arXiv:2009.03774 [gr-qc]].

\bibitem{Doneva:2021dqn}
D.~D.~Doneva and S.~S.~Yazadjiev,
[arXiv:2101.03514 [gr-qc]].

\bibitem{Herdeiro:2018wub}
C.~A.~R.~Herdeiro, E.~Radu, N.~Sanchis-Gual and J.~A.~Font,
Phys. Rev. Lett. \textbf{121}, no.10, 101102 (2018)

\bibitem{Myung:2018vug}
Y.~S.~Myung and D.~C.~Zou,
Eur. Phys. J. C \textbf{79}, no.3, 273 (2019)

\bibitem{Boskovic:2018lkj}
M.~Boskovic, R.~Brito, V.~Cardoso, T.~Ikeda and H.~Witek,
Phys. Rev. D \textbf{99}, no.3, 035006 (2019)

\bibitem{Myung:2018jvi}
Y.~S.~Myung and D.~Zou,
Phys.\ Lett.\ B \textbf{790}, 400-407 (2019)

\bibitem{Fernandes:2019rez}
P.~G.~S.~Fernandes, C.~A.~R.~Herdeiro, A.~M.~Pombo, E.~Radu and N.~Sanchis-Gual,
Class. Quant. Grav. \textbf{36}, no.13, 134002 (2019)
[erratum: Class. Quant. Grav. \textbf{37}, no.4, 049501 (2020)]

\bibitem{Brihaye:2019kvj}
Y.~Brihaye and B.~Hartmann,
Phys. Lett. B \textbf{792}, 244-250 (2019)

\bibitem{Myung:2019oua}
Y.~S.~Myung and D.~C.~Zou,
Eur. Phys. J. C \textbf{79}, no.8, 641 (2019)

\bibitem{Astefanesei:2019pfq}
D.~Astefanesei, C.~Herdeiro, A.~Pombo and E.~Radu,
JHEP \textbf{10}, 078 (2019)

\bibitem{Konoplya:2019goy}
R.~A.~Konoplya and A.~Zhidenko,
Phys. Rev. D \textbf{100}, no.4, 044015 (2019)

\bibitem{Fernandes:2019kmh}
P.~G.~S.~Fernandes, C.~A.~R.~Herdeiro, A.~M.~Pombo, E.~Radu and N.~Sanchis-Gual,
Phys. Rev. D \textbf{100}, no.8, 084045 (2019)

\bibitem{Zou:2019bpt}
D.~C.~Zou and Y.~S.~Myung,
Phys. Rev. D \textbf{100}, no.12, 124055 (2019)

\bibitem{Brihaye:2019gla}
Y.~Brihaye, C.~Herdeiro and E.~Radu,
Phys. Lett. B \textbf{802}, 135269 (2020)

\bibitem{Astefanesei:2019qsg}
D.~Astefanesei, J.~L.~Bl\'azquez-Salcedo, C.~Herdeiro, E.~Radu and N.~Sanchis-Gual,
JHEP \textbf{07}, 063 (2020)

\bibitem{Blazquez-Salcedo:2020nhs}
J.~L.~Bl\'azquez-Salcedo, C.~A.~R.~Herdeiro, J.~Kunz, A.~M.~Pombo and E.~Radu,
Phys. Lett. B \textbf{806}, 135493 (2020)

\bibitem{Blazquez-Salcedo:2020jee}
J.~L.~Bl\'azquez-Salcedo, C.~A.~R.~Herdeiro, S.~Kahlen, J.~Kunz, A.~M.~Pombo and E.~Radu,
[arXiv:2008.11744 [gr-qc]].

\bibitem{Blazquez-Salcedo:2020crd}
J.~L.~Bl\'azquez-Salcedo, S.~Kahlen and J.~Kunz,
Symmetry \textbf{12}, no.12, 2057 (2020)



\bibitem{Ramazanoglu:2017xbl}
F.~M.~Ramazano\u{g}lu,
Phys. Rev. D \textbf{96}, no.6, 064009 (2017)

\bibitem{Ramazanoglu:2018tig}
F.~M.~Ramazano\u{g}lu,
Phys. Rev. D \textbf{98}, no.4, 044013 (2018)

\bibitem{Ramazanoglu:2019gbz}
F.~M.~Ramazano\u{g}lu,
Phys. Rev. D \textbf{99}, no.8, 084015 (2019)

\bibitem{Ramazanoglu:2019jrr}
F.~M.~Ramazano\u{g}lu and K.~\.I.~\"Unl\"ut\"urk,
Phys. Rev. D \textbf{100}, no.8, 084026 (2019)


\bibitem{Fan:2016jnz}
Z.~Y.~Fan,
JHEP \textbf{09}, 039 (2016)


\bibitem{Oliveira:2020dru}
J.~M.~S.~Oliveira and A.~M.~Pombo,
[arXiv:2012.07869 [gr-qc]].





\bibitem{Wald:1984rg}
R.~M.~Wald,
``General Relativity,''
(Chicago Univ.~Pr., Chicago, USA, 1984).

\bibitem{Lee:1990nz}
J.~Lee and R.~M.~Wald,
J. Math. Phys. \textbf{31}, 725-743 (1990)

\bibitem{Wald:1993nt}
R.~M.~Wald,
Phys. Rev. D \textbf{48}, no.8, 3427-3431 (1993)

\bibitem{Iyer:1994ys}
V.~Iyer and R.~M.~Wald,
Phys. Rev. D \textbf{50}, 846-864 (1994)

\bibitem{Hajian:2015xlp}
K.~Hajian and M.~M.~Sheikh-Jabbari,
Phys. Rev. D \textbf{93}, no.4, 044074 (2016)

\bibitem{Ghodrati:2016vvf}
M.~Ghodrati, K.~Hajian and M.~R.~Setare,
Eur. Phys. J. C \textbf{76}, no.12, 701 (2016)

\bibitem{Hajian:2020dcq}
K.~Hajian, S.~Liberati, M.~M.~Sheikh-Jabbari and M.~H.~Vahidinia,
Phys. Lett. B \textbf{812}, 136002 (2020)


\bibitem{Ascher:1979iha}
U.~Ascher, J.~Christiansen and R.~D.~Russell,
Math. Comput. \textbf{33}, no.146, 659-679 (1979)

\end{thebibliography}
\end{document}